\documentclass[]{JHEP3}%
\usepackage{graphics}      
\usepackage{epsfig,latexsym}

\newcommand{\be}{\begin{equation}}
\newcommand{\ee}{\end{equation}}
\newcommand{\bea}{\begin{eqnarray}}
\newcommand{\eea}{\end{eqnarray}}
\newcommand{\lsim}{\mbox{\raisebox{-.6ex}{~$\stackrel{<}{\sim}$~}}}
\newcommand{\gsim}{\mbox{\raisebox{-.6ex}{~$\stackrel{>}{\sim}$~}}}

\newcommand{\sH}{\mathcal{H}}

\def\grad{\vec{\nabla}}

\title{Phenomenology of a Pseudo-Scalar Inflaton: Naturally Large Nongaussianity}

\author{Neil Barnaby, Ryo Namba, Marco Peloso\\
School of Physics and Astronomy, University of Minnesota, Minneapolis, MN 55455, USA\\
E-mail: \email{barnaby@physics.umn.edu},  \email{namba@physics.umn.edu}, \email{peloso@physics.umn.edu}}

\preprint{}

\abstract{
Many controlled realizations of chaotic inflation employ pseudo-scalar axions. Pseudo-scalars $\phi$ are naturally coupled to gauge  
fields through $c\phi F \tilde{F}$.  In the presence of this coupling, gauge field quanta are copiously produced by the rolling inflaton.
The produced gauge quanta, in turn, source inflaton fluctuations via inverse decay.  These new cosmological perturbations add incoherently with the ``vacuum''
perturbations, and are highly nongaussian.  This provides a natural mechanism to generate large nongaussianity in single or multi field slow-roll inflation. The  
resulting phenomenological signatures are highly distinctive: large  nongaussianity of (nearly) equilateral shape, in addition to detectably large values of both
the scalar spectral tilt and tensor-to-scalar ratio (both being typical of large field inflation).  The WMAP bound on nongaussianity implies that the coupling $c$ 
of the pseudo-scalar inflaton to any gauge field must be smaller than about $10^{2}\,M_p^{-1}$.
}

\begin{document}

\section{Introduction}
\label{sec:intro}


Primordial inflation provides a simple mechanism to resolve the conceptual difficulties of the standard Big Bang
cosmology and has enjoyed great phenomenological success in accounting for the properties of the observed 
Cosmic Microwave Background (CMB) anisotropies.  As such, inflation has become the dominant paradigm for  the early universe.  
In spite of this success, however, a compelling particle physics realization is still lacking. 
The key obstruction is the requirement of a suitably  flat scalar potential, $V(\varphi)$.  Successful inflation requires 
$\epsilon,|\eta|\ll 1$ where the slow roll parameters are defined as
\begin{equation}
  \epsilon\equiv \frac{M_p^2}{2}\left(\frac{V'}{V}\right)^2, \hspace{5mm} \eta \equiv M_p^2\frac{V''}{V} \label{SR}
\end{equation}
Here prime denotes derivative with respect to $\varphi$, and $M_p \simeq 2.4 \cdot 10^{18} \, {\rm GeV}$ is the reduced Plank mass. These parameters are notoriously sensitive to Ultra-Violet (UV) 
physics.  For example, even generic Planck-suppressed corrections to $V(\varphi)$ may contribute $\Delta \eta = \mathcal{O}(1)$, thus spoiling inflation.  This UV sensitivity represents a technical fine
tuning problem which must be addressed in any particle physics model of inflation.

It is conceivable that dangerous corrections to $\epsilon,\eta$ may be absent as a result of fine-tuning \cite{delicate}, or that the requirement of a flat potential 
can be evaded by invoking somewhat exotic effects such as dissipation \cite{warm,trapped}, small sound speed \cite{DBI} or higher derivative corrections \cite{NL}.  However, perhaps the 
simplest and most cogent way to realize $\epsilon,|\eta|\ll 1$ in a natural way is by assuming that the inflaton $\varphi$ is a Pseudo-Nambu-Goldstone-Boson (PNGB) 
\cite{natural,natural1.5,extranatural,2-flation, N-flation,N-flation2,monodromy,monodromy2,kaloper,lorenzo}.  In this case the inflaton enjoys a shift symmetry $\varphi \rightarrow \varphi + \mathrm{const}$, 
which is broken either explicitly or by quantum effects.  In the limit of exact symmetry we must have $\epsilon=\eta=0$, thus dangerous contributions to the slow roll parameters 
are controlled by the smallness of symmetry breaking.  Moreover, PNGBs like the axion are ubiquitous in particle physics: they
arise whenever an approximate global symmetry is spontaneously broken and are plentiful in string theory compactifications.

The idea of invoking a PNGB to obtain a natural realization of inflation is more than 20 years old.  The first model, natural inflation 
\cite{natural}, exploited the periodic potential
\begin{equation}
\label{periodic}
  V(\varphi) = \Lambda^4\left[ 1 - \cos\left(\frac{\varphi}{f}\right) \right]
\end{equation}
which arises from nonperturbative effects and breaks the continuous shift symmetry down to a discrete subgroup $\varphi\rightarrow\varphi + (2\pi)f$.  Unfortunately, this model is compatible with observation 
only when the axion decay constant is $f > M_p$ \cite{natural2}, a regime that may be impossible to realize in a controlled effective field 
theory because it suggests a global symmetry broken \emph{above} the quantum gravity 
scale \cite{extranatural}.  It has also been conjectured that $f>M_p$ cannot be realized in a controlled limit of string theory \cite{big_f}.  Fortunately, this difficulty can be easily evaded, for example by 
considering two \cite{2-flation} or more \cite{N-flation,N-flation2} axion fields, extra dimensions \cite{extranatural}, or by 
exploiting the non-periodic contributions to $V(\varphi)$ that arise from wrapping branes on suitable cycles \cite{monodromy,monodromy2}.  Currently, there exist a number of natural, 
controlled realizations of axion inflation with an axion scale $f$ a few orders of magnitude smaller 
than $M_p$ but that  nevertheless behave effectively as large field inflation models \cite{2-flation,N-flation,N-flation2,monodromy,monodromy2,kaloper,lorenzo}.

In any axion inflation model, the inflaton is expected to couple to some gauge field $A_\mu$ via interactions of the type
\begin{equation}
\label{int}
  \mathcal{L}_{\mathrm{int}} = - \frac{\alpha}{4 f}\, \varphi \, F^{\mu\nu}\tilde{F}_{\mu\nu}
\end{equation}
where $F_{\mu\nu} = \partial_\mu A_\nu - \partial_\nu A_\mu$ is the field strength and $\tilde{F}^{\mu\nu} = \frac{1}{2}\epsilon^{\mu\nu\alpha\beta} F_{\alpha\beta}$ is its dual.  The strength of the interaction is 
controlled by the decay constant, $f$, and by the dimensionless parameter $\alpha$.  While $\alpha$ is in principle a model dependent quantity, from the perspective of effective field theory we generally expect it
to be order unity.  On the
other hand,  Ref.~\cite{lorenzo} has provided concrete examples (in multi-field or extra-dimensional models) which can result in $\alpha$ greater or much greater than one.  Thus, for controlled effective field
theory realizations of axion inflation it is very natural to have $\alpha / f \gg M_p^{-1}$, in which case the interaction (\ref{int}) 
is much stronger than gravitationally suppressed.  In this work, we consider in detail the cosmological implications of the interaction 
(\ref{int}), which has been neglected in nearly all previous studies.  Our analysis is quite general: the general logic of effective field
theory requires the inclusion of an interaction (\ref{int}) whenever $\phi$ is pseudo-scalar.

In \cite{lorenzo} it was shown that energy dissipation into gauge fields, via the interaction (\ref{int}), can slow the motion of the 
inflaton on a very steep potential.  A  more conservative approach was adopted in our recent work \cite{ai}, where we note that 
the interaction (\ref{int}) can have a profound impact on the phenomenology of the model, even in the conventional slow roll regime.  
The underlying physics is as follows. The motion of the inflaton amplifies the fluctuations of the gauge field, $\delta A$, which in turn 
produce inflaton fluctuations via \emph{inverse decay}: $\delta A + \delta A \rightarrow \delta \varphi$.  
When $f /\alpha \lsim 10^{-2} M_p$, which is very natural in models that admit a UV completion, this new source of perturbations actually 
dominates over the usual fluctuations from the vacuum.  In this regime, all 
previous studies of axion inflation are invalid.  Our analysis is phenomenological: we use CMB data to place observational limits on the 
coupling $\alpha/f$ without making any specific assumptions about the microphysical origins of the model.  We believe that this study should 
serve as motivation for a case-by-case analysis of the allowed values of $\alpha$ in various explicit realizations of axion inflation.

In this paper, we reconsider the cosmological fluctuations in axion inflation, extending significantly our previous work \cite{ai}. We are motivated, in part, by the recent surge of interest in computing and 
measuring nongaussian effects in the CMB (see \cite{NGreview} for a recent review).  Nongaussian statistics, such as the bispectrum, provide a powerful tool to discriminate between the plethora of inflationary 
models in the literature and may provide a valuable window into the detailed physics of the very early universe.  However, a single decoupled scalar field in slow roll is well known to produce an undetectably 
small signal \cite{riotto,maldacena,seerylidsey}.  The reason for this is intuitively easy to understand: nongaussianity is a measure of the strength of interactions, while in the vanilla scenario the requirement 
of a flat potential typically also constrains interactions to be weak.  To evade this no-go result and obtain an observably large signal, previous studies have invoked non-standard field theories (with small sound 
speed \cite{small_sound} or higher derivatives \cite{NL}) or initial conditions \cite{small_sound,nonBD1,nonBD2,nonBD}, potentials with sharp features \cite{chen1,chen2}, dissipative effects \cite{trapped}, 
fine-tuned inflationary trajectories \cite{turnNG} or post-inflationary effects (such as preheating \cite{preheatNG,preheatNG2}).  In the work \cite{ai}, on the other hand, the no-go results of 
\cite{riotto,maldacena,seerylidsey} are circumvented in a very novel way: the interaction that gives rise to large nongaussianity, eqn.~(\ref{int}), does not play any role in the background dynamics and is thus 
unconstrained by the requirement of slow roll.  

In terms of the broader picture we believe that this work -- along with \cite{ai} and the previous studies \cite{pp1,pp2,pp3} -- suggests that the conventional lore concerning the difficulty of obtaining large 
nongaussianity may have been excessively conservative.  One generically expects that inflaton to couple to some fields which do not play any role in driving inflation.  Such interactions are unavoidable from an
effective field theory perspective and are (at least to some extent) necessary in order to successfully reheat the universe after inflation.  Refs. \cite{ai,pp1,pp2,pp3}  provide explicit examples demonstrating how the consistent
inclusion of such interactions can radically modify the phenomenology of inflation, via particle production effects.  This work represents a challenge to the conventional lore that nongaussianity is a ``smoking gun'' 
signature of non-standard inflationary dynamics by illustrating explicitly that perhaps the simplest and best-motivated particle 
physics models of inflation are \emph{already} constrained by existing observational limits on nongaussianity.

This paper is organized as follows. In section \ref{sec:overview} we provide an overview of the mechanism and briefly summarize our previous work \cite{ai}. In section \ref{sec:correlators} we compute the correlation 
functions of the curvature fluctuations and tensor (gravity wave, GW) perturbations.  Section \ref{sec:pheno} studies the resulting phenomenology. We have written this section in self-contained way, so that a reader 
who is interested only in the observational predictions can skip the previous more technical section. In section \ref{sec:perts} we perform a complete computation of the perturbations, that includes also the metric 
perturbations. We show that metric perturbations can be neglected (at leading order), in the computation of the density and GW correlators sourced by the gauge field. In section \ref{sec:models} we review most of the 
existing models of axion inflation, and briefly discuss the implications of our findings for these models. Finally, in section \ref{sec:conclusions}, we conclude.

\section{Overview of the Mechanism}
\label{sec:overview}

In this section we provide a brief overview of the production of gauge field fluctuations and the subsequent inverse decay effects in axion inflation.  
This section is largely review of \cite{ai}.

We consider a simple theory of a PNGB inflaton interacting with a $U(1)$ gauge field\footnote{The generalization to non-Abelian gauge groups is straightforward.} via the interaction (\ref{int}).  The action is
\begin{equation}
\label{L}
  \mathcal{S} =  \int d^4 x \sqrt{-g} \left[ \frac{M_p^2}{2} \, R  -\frac{1}{2}(\partial \varphi)^2 - V(\varphi) - \frac{1}{4}F^{\mu\nu}F_{\mu\nu} - \frac{\alpha}{4 f} \varphi  \, \tilde{F}^{\mu\nu} \, F_{\mu\nu} \right]
\end{equation}
where $R$ is the Ricci scalar, $F_{\mu \nu} = \partial_\mu A_\nu - \partial_\nu A_\mu$ the field strength, and ${\tilde F}^{\mu \nu} \equiv \frac{1}{2} \, \frac{\eta^{\mu \nu \alpha \beta}}{\sqrt{-g}} \, F_{\alpha \beta}$
its dual, with $\eta^{0123} = 1$. We separate the inflaton into a homogeneous (background) part plus its fluctuations
\begin{equation}
  \varphi = \phi \left( t \right) + \delta \varphi \left( t ,\, \vec{x} \right)
\end{equation}
We leave the potential $V(\varphi)$ arbitrary, except to assume that it is sufficiently flat to support the required amount of inflation ($N_e \gsim 60$).  We assume a spatially flat Friedmann-Robertson-Walker 
(FRW) space-time with metric
\begin{eqnarray}
  ds^2 \equiv g_{\mu\nu} dx^{\mu} dx^{\nu} &=& -dt^2 + a^2(t)\, d{\bf x} \cdot d{\bf x} \\
  &=& a^2(\tau) \left[ -d\tau^2 +  d{\bf x} \cdot d{\bf x}  \right]
\end{eqnarray}
where on the second line we have introduced conformal time, $\tau$, related to cosmic time as $ad\tau = dt$.  Derivatives with respect to
cosmic time are denoted as $\partial_t f \equiv \dot{f}$ and with respect to conformal time as $\partial_\tau f \equiv f'$.  The Hubble rate $H \equiv \dot{a} / a$
has conformal time analogue $\sH \equiv a' / a$.  

We are first interested in the gauge quanta which are produced by the homogeneous rolling inflaton $\phi(\tau)$. To this end, we can ignore the inflaton and metric perturbations in the equations of motion of the 
gauge field (see section \ref{sec:perts} for the complete treatment). Extremizing the action with respect to $A_0$, and choosing  the Coulomb gauge $A_0 =  0$, then gives $\left( \grad \cdot \vec{A} \right)' = 0$, 
from which we set $\grad \cdot \vec{A} = 0$. The equations of motion for $\vec{A}$ then read
\begin{equation}
\vec{A}'' - \nabla^2 \vec{A} - \frac{\alpha}{f} \, \phi' \, \grad \times \vec{A} = 0
\label{eqA}
\end{equation}
As we discuss in subsection \ref{subsec:production}, this equation describes the production of the quanta of the gauge fields that results from the motion of the inflaton.  

The produced gauge quanta have two key
effects: they backreact on the homogeneous background dynamics (see subsection \ref{subsec:backreaction}) and also source inflaton perturbations (see subsection \ref{subsec:id}).  Both effects are governed by the 
equation of motion of the inflaton, and the $00$ Einstein equation, which read, respectively
\begin{eqnarray}
&& \varphi'' + 2 {\cal H} \varphi' - \nabla^2 \varphi  + a^2 \, \frac{d V}{d \varphi}  = a^2 \frac{\alpha}{f}  \vec{E}\cdot \vec{B} \nonumber\\
&& {\cal H}^2 = \frac{1}{3 M_p^2} \left[ \frac{1}{2} \varphi'^2 + \frac{1}{2} \left(  \grad \varphi \right)^2   + a^2 \, V + \frac{a^2}{2} \left( \vec{E}^2 + \vec{B}^2 \right) \right]
\label{eqphi-00}
\end{eqnarray}
In these equations we have retained the spatial dependence of $\varphi$ (due to the inflaton perturbations), and we have introduced  the physical ``electric'' and ``magnetic'' 
fields\footnote{We do not assume that $A_\mu$ necessarily corresponds to the Standard Model electro-magnetic gauge potential.} 
\begin{equation}
\label{electromagnetic}
  \vec{B} = \frac{1}{a^2} \grad\times\vec{A},\hspace{5mm} \vec{E} = -\frac{1}{a^2} \vec{A}'
\end{equation}

\subsection{Production of Gauge Field Fluctuations}
\label{subsec:production}

During inflation, the motion of the inflaton leads to an instability for the fluctuations of the gauge field.  To see this effect, we start from the equation of motion for $A_\mu$ in the background of the homogeneous 
inflaton $\phi(t)$, eq. (\ref{eqA}) above.  We decompose the q-number field $\vec{A}(\tau,{\bf x})$ as
\begin{equation}
  \vec{A}(\tau,{\bf x}) = \sum_{\lambda=\pm} \int \frac{d^3k}{(2\pi)^{3/2}} \left[ \vec{\epsilon}_\lambda({\bf k}) a_{\lambda}({\bf k}) A_\lambda(\tau,{\bf k}) e^{i {\bf k}\cdot {\bf x}} + \mathrm{h.c.}   \right]
\label{decomposition}
\end{equation}
where ``$\mathrm{h.c.}$'' denotes the Hermitian conjugate of the preceding term, the annihilation/creation operators obey
\begin{equation}
  \left[a_{\lambda}({\bf k}), a_{\lambda'}^{\dagger}({\bf k'})\right] = \delta_{\lambda\lambda'}\delta^{(3)}({\bf k}-{\bf k'})
\label{ladder}
\end{equation}
Here $\vec{\epsilon}_\lambda$ are circular polarization vectors satisfying  $\vec{k}\cdot \vec{\epsilon}_{\pm} \left( \vec{k} \right) = 0$, 
$\vec{k} \times \vec{\epsilon}_{\pm} \left( \vec{k} \right) = \mp i k \vec{\epsilon}_{\pm} \left( \vec{k} \right)$,
$\vec{\epsilon}_\pm \left( \vec{-k} \right) = \vec{\epsilon}_\pm \left( \vec{k} \right)^*$, and normalized according to $\vec{\epsilon}_\lambda \left( \vec{k} \right)^* 
\cdot \vec{\epsilon}_{\lambda'} \left( \vec{k} \right) = \delta_{\lambda \lambda'}$.  

Inserting the decomposition (\ref{decomposition}) into eq. (\ref{eqA}) results in the equation of motion
\begin{equation}
\label{Amode}
  \left[ \frac{\partial^2}{\partial\tau^2} + k^2 \pm \frac{2 k \xi}{\tau} \right] A_{\pm}(\tau,k) = 0, \hspace{5mm} \xi \equiv \frac{\alpha \dot{\phi}}{2 f H}
\end{equation}
for the c-number mode functions $A_\pm$. During inflation the parameter $\xi$ may be treated as constant, as its time variation is subleading in a slow roll expansion.

From equation (\ref{Amode}) we see that one of the polarizations of $\vec{A}_{\lambda}$ experiences a tachyonic instability for $k/(aH) \lsim 2\xi$. Without loss of generality, 
we assume that $\dot{\phi} > 0$ during inflation, so that the mode exhibiting the instability is $A_+$. In appendix A we review the solutions of (\ref{Amode}) and show that the 
growth of fluctuations is well described by \cite{lorenzo}
\begin{equation}
\label{mode_soln}
  A_{+}(\tau,k) \cong \frac{1}{\sqrt{2k}} \left(\frac{k}{2\xi a H}\right)^{1/4} e^{\pi \xi - 2\sqrt{2 \xi k / (aH)}}
\end{equation}
in the interval $(8\xi)^{-1} \lsim k/(aH) \lsim 2\xi$ \cite{ai} of phase space that accounts for most of the power in the produced gauge fluctuations.  The phase space of growing modes is 
non-vanishing for $\xi \gsim \mathcal{O}(1)$, which we assume throughout.  Notice the exponential enhancement $e^{\pi \xi}$ in the solution (\ref{mode_soln}), which arises due to tachyonic
instability, and reflects significant nonperturbative gauge particle production in the regime $\xi \gsim 1$.  On the other hand, the production of gauge field fluctuations is uninterestingly small
for $\xi < 1$.  Note also that the other polarization state, $A_{-}(\tau,k)$, is not produced and can therefore be ignored.

We have thus seen that the motion of the homogeneous inflaton $\phi(t)$ leads to production of gauge field quanta $\delta A_\mu$.  There are two key physical 
effects associated with the interactions of these produced quanta with the inflaton.  The first effect is the backreaction of the produced quanta on the 
homogeneous dynamics of $\phi(t)$, $a(t)$.  In the next subsection we study the conditions under which backreaction effects are negligible.  The
second key physical effect is the production of inflaton fluctuations via \emph{inverse decay}; this is the subject of subsection \ref{subsec:id}.

\subsection{Backreaction Effects}
\label{subsec:backreaction}

Backreaction effects can be accounted for using the mean of the field equations (\ref{eqphi-00}):
\begin{eqnarray}
  &&  \ddot{\phi} + 3 H \dot{\phi} + V'(\phi) = \frac{\alpha}{f} \langle \vec{E}\cdot \vec{B} \rangle \label{mean1} \\
  && 3 H^2 = \frac{1}{M_p^2} \left[ \frac{1}{2}\dot{\phi}^2 + V(\phi) + \frac{1}{2} \langle \vec{E}^2 + \vec{B}^2 \rangle \right] \label{mean2}
\end{eqnarray}
where we have switched to physical time. The expectation values appearing in (\ref{mean1},\ref{mean2}) encode the backreaction of the produced gauge quanta on the homogeneous
dynamics of $\phi(t)$, $a(t)$.  From (\ref{electromagnetic}) and (\ref{decomposition}), we have
\begin{eqnarray}
\langle \vec{E} \cdot \vec{B} \rangle &=& - \frac{1}{4 \pi^2 a^4} \int d k \, k^3 \, \frac{d}{d \tau} \vert A_+ \vert^2 \nonumber\\
\frac{1}{2}\langle \vec{E}^2+\vec{B}^2 \rangle &=& \frac{1}{4 \pi^2 a^4} \int d k \,  k^2 \left[ \vert A_+' \vert^2 + k^2 \vert A_+ \vert^2 \right]
\label{integrals-mean}
\end{eqnarray}
Since we are studying the backreaction of the produced quanta, we should disregard the modes that do not experience the growth discussed 
in the previous subsection, namely all modes $A_-$ and the large momentum ($k > 2\xi a H$) modes of $A_+$.  As can be seen from (\ref{Amode}), 
modes of $A_{+}$ with $k/(aH) \gg 2\xi$ remain in their vacuum state and do not experience any tachyonic instability.
Such modes contribute to the vacuum energy of the $U(1)$ field (in eqn.~(\ref{mean2}); an analogous consideration applies to
$\langle \vec{E} \cdot \vec{B} \rangle$ in eqn.~(\ref{mean1})) that we assume is canceled by a bare vacuum energy, as is customary
in QFT. (In short, we have nothing to add to the cosmological constant problem.)  This prescription provides a UV cutoff 
$k / \left( a H \right) < 2 \xi$ in the integrals (\ref{integrals-mean}); see the paragraph after eq. (\ref{Amode}).  From a direct 
inspection of the solutions of (\ref{Amode}), we verified that the integrals (\ref{integrals-mean}) converge in the infrared 
$k \rightarrow 0$ region, and that they receive their support almost entirely from the region $(8\xi)^{-1} \lsim k/(aH) \lsim 2\xi$ in which 
(\ref{mode_soln}) is a very good approximation of the exact solution. Therefore, (\ref{integrals-mean}) can be evaluated by using the expression  (\ref{mode_soln}) for $A_+$, and by  integrating  
only over this region of momenta. In fact, we verified that the momentum interval can be extended from $0$ to $\infty$, since  the expression (\ref{mode_soln}) rapidly decreases outside the 
$(8\xi)^{-1} \lsim k/(aH) \lsim 2\xi$ interval, and the contribution of the ``outer'' regions to (\ref{integrals-mean}) can be neglected. Proceeding in this way allows for an analytic result:
\begin{equation}
\langle \vec{E} \cdot \vec{B} \rangle \simeq - 2.4 \cdot 10^{-4} \, \frac{H^4}{\xi^4} \, {\rm e}^{2 \pi \xi}
\;\;\;,\;\;\;
\langle \frac{\vec{E}^2+\vec{B}^2}{2} \rangle \simeq 1.4 \cdot 10^{-4} \, \frac{H^4}{\xi^3} \, {\rm e}^{2 \pi \xi}
\label{result-mean}
\end{equation}
This procedure follows Ref.~\cite{lorenzo}, where these expressions were first derived. We expect that any sensible renormalization prescription will yield results in agreement with (\ref{result-mean}).

From (\ref{mean1},\ref{mean2}) we can distinguish two distinct kinds of backreaction effects.  First, the gauge field fluctuations are produced at the expense of the kinetic energy of $\phi(t)$.  This 
contributes a new source of dissipation into the homogeneous Klein-Gordon equation (\ref{mean1}).  In order to trust the usual slow-roll inflationary solution, we require that $ \vert 3H \dot{\phi} \vert 
\cong \vert  -V'(\phi) \vert \gg  \vert \frac{\alpha}{f}\langle \vec{E}\cdot \vec{B} \rangle \vert $.  Using (\ref{result-mean}), this condition reads
\begin{equation}
\label{back1}
  \frac{H^2}{2\pi |\dot{\phi}|} \ll 13\, \xi^{3/2} e^{-\pi\xi}
\end{equation}

The second kind of backreaction effect arises because the energy density in produced gauge field fluctuations contributes to the Friedmann equation 
(\ref{mean2}).   To ensure that the expansion of the universe is dominated by the potential energy of the inflaton we require $3 M_p^2 H^2 \cong V \gg \frac{1}{2} \langle \vec{E}^2 + \vec{B}^2 \rangle$. Once eq. (\ref{result-mean}) is taken into account, this condition reads
\begin{equation}
\label{back2}
  \frac{H}{M_p} \ll 146\, \xi^{3/2} e^{-\pi\xi}
\end{equation}
Taken together, the constraints (\ref{back1},\ref{back2}) ensure that the unstable growth of gauge field fluctuations does not modify the usual 
homogeneous inflationary dynamics.  

\subsection{Inverse Decay Effects and Inflaton Perturbations}
\label{subsec:id}

Even when (\ref{back1},\ref{back2}) are satisfied, the coupling $\varphi F \tilde{F}$ may still have a profound impact on the cosmological fluctuations in the model (\ref{L}).  The perturbations of the inflaton are 
described by the equation
\begin{equation}
\label{phi_eqn}
  \left[ \frac{\partial^2}{\partial\tau^2} +2 \sH \frac{\partial}{\partial\tau} - \nabla^2 + a^2 m^2 \right] \delta\varphi(\tau,{\bf x}) 
  = a^2 \frac{\alpha}{f} \, \left(  \vec{E}\cdot\vec{B} - \langle \vec{E}\cdot\vec{B} \rangle \right)
\end{equation}
where $m^2 \equiv V''$. The solution of (\ref{phi_eqn}) splits into two parts: the solution of the homogeneous equation and the particular solution which is due to the source.  Schematically, we have
\begin{equation}
  \delta\varphi(\tau,{\bf x}) = \underbrace{\delta\varphi_{\mathrm{vac}}(\tau,{\bf x})}_{\mathrm{homogeneous}} + \underbrace{\delta\varphi_{\mathrm{inv.decay}}(\tau,{\bf x})}_{\mathrm{particular}}
\label{hom-inhom}
\end{equation}
The homogeneous solution corresponds, physically, to the usual vacuum fluctuations from inflation.  The particular solution, on the other hand, can be 
interpreted as arising due to inverse decay processes $\delta A + \delta A \rightarrow \delta \varphi$.  This new source of inflaton fluctuations contributes
directly to the observable curvature perturbation on uniform density hypersurfaces, owing to the relation $\zeta \sim - \frac{H}{\dot{\phi}}\delta \varphi$. It was shown in \cite{ai} that the inverse decay contribution to the cosmological fluctuations may actually \emph{dominate} over the usual vacuum fluctuations
in the regime $f \lsim 10^{-2} M_p$.  This radically modifies the phenomenology of axion inflation.  In particular, the inverse decay contribution to the primordial
cosmological fluctuations is highly nongaussian; this is evident already from (\ref{phi_eqn}) since the particular solution $\delta\varphi_{\mathrm{inv.decay}}$ is bilinear 
in the gaussian field $\delta A_\mu$.  

To proceed with the computation, we decompose
\begin{equation}
\label{fourier}
  \delta\varphi(\tau,{\bf x}) = \int \frac{d^3k}{(2\pi)^{3/2}} \frac{Q_{\bf k}(\tau)}{a(\tau)} e^{i {\bf k}\cdot {\bf x}}
\end{equation}
Eq. (\ref{phi_eqn}) then results in (we note that the last term of (\ref{phi_eqn}) has  no effect on the mode functions with momentum 
different from zero) \footnote{In Section \ref{sec:perts} we present the complete computation, including also scalar metric perturbations. 
We show explicitly that equation (\ref{phi_cor}) still holds, with only the addition of subdominant (Planck-suppressed) terms in the source 
$J_{\bf k}$.}
\begin{eqnarray}
&&  \left[\partial_\tau^2 + k^2 + a^2 m^2 - \frac{a''}{a} \right] Q_{\bf k}(\tau) = J_{\bf k}(\tau) \label{phi_cor}\\
&& J_{\bf k}(\tau) \equiv a^3(\tau) \frac{\alpha}{f} \int \frac{d^3k}{(2\pi)^{3/2}} e^{-i{\bf k}\cdot {\bf x}} \vec{E}\cdot{\vec{B}}  
\label{source}
\end{eqnarray}
We separate the mode functions of the two terms in eq. (\ref{hom-inhom}) as
\begin{equation}
  Q_{\bf k}(\tau) = Q_{\bf k}^{\mathrm{vac}}(\tau) + Q_{\bf k}^{\mathrm{inv.decay}}(\tau)
\label{hom-inhom2}
\end{equation}
and we discuss each contribution separately.

The homogeneous term is expanded  as
\begin{equation}
\label{hom}
  Q_{\bf k}^{\mathrm{vac}}(\tau) = b({\bf k}) \varphi_k(\tau) + b^\dagger({-{\bf k}}) \varphi_k^\star(\tau)
\end{equation}
The inflaton ladder operators obey
\begin{equation}
  \left[b({\bf k}), b^\dagger({\bf k'})\right] = \delta^{(3)}({\bf k}-{\bf k'})
\label{ladder-phi}
\end{equation}
and commute with the ladder operators of the gauge field
\begin{equation}
   \left[b({\bf k}), a_\lambda({\bf k'})\right] = \left[b({\bf k}), a^\dagger_\lambda({\bf k'})\right] = 0
\end{equation}
The properly normalized homogeneous solutions of (\ref{phi_cor}) are given by the well-known result
\begin{equation}
\label{free_mode}
  \varphi_k(\tau) = i \frac{\sqrt{\pi}}{2}\sqrt{-\tau} H^{(1)}_{\nu}(-k\tau), \hspace{5mm}\nu \cong \frac{3}{2} + \mathcal{O}(\epsilon,\eta)
\end{equation}
where we have chosen the (arbitrary) phase so that $\varphi_k(\tau)$ is real in the limit $-k\tau\rightarrow 0$.   

The vacuum modes (\ref{free_mode}) are employed in the retarded Green function associated with (\ref{phi_cor}),
\begin{equation}
\label{green}
  G_k(\tau,\tau') = i \Theta(\tau-\tau') \left[ \varphi_k(\tau)\varphi_k^\star(\tau') - \varphi_k^\star(\tau)\varphi_k(\tau') \right]
\end{equation}
which obeys $ \left[\partial_\tau^2 + k^2 + a^2 m^2 - \frac{a''}{a} \right]   G_k(\tau,\tau') = \delta \left( \tau - \tau' \right)$. 

Using the Green function (\ref{green}) the particular solution of (\ref{phi_cor}) takes the form
\begin{equation}
\label{par}
  Q^{\mathrm{inv.decay}}_{\bf k}(\tau) = \int_{-\infty}^{0}d\tau' G_k(\tau,\tau') J_{\bf k}(\tau')
\end{equation}
where the source term was defined in eq. (\ref{source}). We note that this particular solution is statistically independent of the homogeneous solution (\ref{hom}).  
In fact, the particular solution can be expanded in terms of the annihilation/creation operators $a_\lambda({\bf k}),a_\lambda^\dagger({\bf k})$ associated with the 
gauge field, while the homogeneous solution can be expanded in terms of the annihilation/creation operators $b({\bf k}),b^\dagger({\bf k})$ associated with the inflaton 
vacuum fluctuations. As we  pointed out,  these two sets of operators commute with one another.

We are now in a position to compute the correlation functions for the perturbations $\delta\varphi$. We present this computation in section \ref{sec:correlators}. In section 
\ref{sec:pheno} we discuss the resulting phenomenology. Finally, in section \ref{sec:perts} we show that these result are valid also when  metric perturbations are also 
consistently taken into account.

\section{Correlation Functions}
\label{sec:correlators}

In this section we compute the main phenomenological signatures of the model (\ref{L}). Specifically: in subsection \ref{sub:2point} we compute the two point correlation function of 
the density perturbation $\zeta = - \frac{H}{\dot{\phi}} \, \delta \varphi$; in subsection \ref{sub:power} we present the corresponding power spectrum; in subsection \ref{sub:3point} 
we compute the three point correlation function of $\zeta$; in subsection \ref{sub:bispectrum} we  present the corresponding bispectrum; in subsection \ref{sub:GW}  we finally summarize 
the computation of the power spectrum of  the gravity waves (GW) modes.  In this section we neglect scalar metric perturbations for simplicity, however, in section \ref{sec:perts} we show
explicitly that their consistent inclusion does not modify our results. 

\subsection{Two-Point Correlation Function}
\label{sub:2point}

We start from the relation $\zeta \left( \tau ,\, \vec{x} \right)  = - \frac{H}{\dot{\phi}} \, \delta \varphi \left( \tau ,\, \vec{x} \right)$ between the inflaton perturbations, and the  
curvature perturbation on uniform density hypersurfaces.\footnote{See section \ref{sec:perts} for more details on this relation.} We decompose the latter as
\begin{equation}
  \zeta (\tau,{\bf x}) = \int \frac{d^3k}{(2\pi)^{3/2}} \, \zeta_{\bf k}(\tau) \,  e^{i {\bf k}\cdot {\bf x}}
\label{zeta-F}
\end{equation}
so that $\zeta_{\bf k} = - \frac{H}{\dot{\phi}} \, \frac{Q_{\bf k}}{a}$.  As we discussed in subsection \ref{subsec:id}, the inflaton perturbations comprise of two terms, one being the vacuum 
fluctuations, and one the fluctuations sourced by the gauge quanta; since these two terms are statistically independent, we have
\begin{equation}
\langle \zeta_{\bf k} \,  \zeta_{\bf k'} \rangle = \frac{H^2}{a^2 \dot{\phi}^2} \left[ \langle 
Q^{\mathrm{vac}}_{\bf k} \, Q^{\mathrm{vac}}_{\bf k'} \rangle   + \langle
Q^{\mathrm{inv.decay}}_{\bf k} \, Q^{\mathrm{inv.decay}}_{\bf k'} \rangle
\right]
\end{equation}
The contribution from the vacuum modes is standard. Using eqs. (\ref{hom}), (\ref{ladder-phi}), and (\ref{free_mode}), one obtains the well-known result
\begin{eqnarray}
  \langle \zeta_{\bf k}^{\mathrm{vac}} \zeta_{\bf k'}^{\mathrm{vac}} \rangle &=&  \frac{H^4}{2 \dot{\phi}^2} \, \left( \frac{k}{a H} \right)^{n_s - 1} \, \frac{1}{k^3} \, \delta^{(3)} \left( {\bf k} + {\bf k}' \right) 
  \nonumber \\
  &=& \frac{2\pi^2}{k^3}\, \mathcal{P} \, \left( \frac{k}{a H} \right)^{n_s - 1}\, \delta^{(3)} \left( {\bf k} + {\bf k}' \right)\label{2vacuum}
\end{eqnarray}
in the late time / large scales limit, $- k \tau \ll 1$.  In the second line of (\ref{2vacuum}) we have introduced
\begin{eqnarray}
  && n_s = 1 + 3-2\nu = 1 + \mathcal{O}(\epsilon,\eta) \\
  && \mathcal{P}^{1/2} \equiv \frac{H^2}{2\pi |\dot{\phi}|} \label{Pcal}
\end{eqnarray}

We now compute the 2-point correlator  of the particular solution (\ref{par}):
\begin{equation}
\langle  \zeta_{\bf k}^{\mathrm{inv.decay}} \left( \tau \right) \, \zeta_{\bf k'}^{\mathrm{inv.decay}} \left( \tau \right)  \rangle
= \frac{H^2}{\dot{\phi}^2} \, \int d \tau' \, d \tau'' \frac{G_k \left( \tau ,\, \tau' \right)}{a(\tau)} \,  \frac{G_{k'} \left( \tau ,\, \tau'' \right)}{a(\tau)} 
\,\,\langle J_{\bf k} \left( \tau' \right) \, J_{\bf k'} \left( \tau'' \right) \rangle
\end{equation}
(In the following, we temporarily omit the superscript `inv.decay' for notational convenience.) In evaluating this expression, we can make some approximations for the 
mode functions $\varphi_k \left( \tau \right)$ and $\varphi_k \left( \tau' \right)$ appearing in the Green function (\ref{green}). Since we are interested in the power 
spectrum of modes well outside the horizon we can use the small argument limit $-k\tau \ll 1$ for the  modes (\ref{free_mode}) entering in the Green function (\ref{green}), 
\begin{equation}
\label{ls_approx}
\varphi_k \left( \tau \right) \simeq \frac{a \left( \tau \right) H}{\sqrt{2}} \, \frac{1}{k^{3/2}} \, \left( - k \tau \right)^{\frac{n_s-1}{2}}, \hspace{5mm} \mathrm{for}-k\tau \ll 1
\end{equation}
Notice that we disregard the small slow roll corrections to the amplitude, but we retain them in the momentum dependence (since this controls the departure of the spectrum from scale invariance). This gives
\begin{equation}
\langle \zeta_{\bf k} \left( \tau \right) \zeta_{\bf k'} \left( \tau \right) \rangle \cong \frac{2 H^4}{\dot{\phi}^2} \, \frac{ \left( - k \tau \right)^{n_s-1} }{k^3}  \int_{-\infty}^\tau d \tau'  \, d \tau''
\, {\rm Im } \left[ \varphi_k \left( \tau' \right) \right]  \, {\rm Im } \left[ \varphi_k \left( \tau'' \right) \right] \, 
\langle J_{\bf k} \left( \tau' \right) \,  J_{\bf k} \left( \tau'' \right) \rangle  \label{cor_step}
\end{equation}
In this relation, we have already used the fact that the correlator is nonvanishing only for $ \vert {\bf k} \vert = \vert {\bf k'} \vert$.  We stress that, in (\ref{cor_step}) we do \emph{not} employ
the approximation (\ref{ls_approx}) for the modes $\varphi_{k}(\tau')$ which appear under the integral, since $\tau'$ must be integrated over.

To evaluate (\ref{cor_step}), we require the source correlator $\langle J_{\bf k} \left( \tau' \right) \,  J_{\bf k} \left( \tau'' \right) \rangle$.  Explicit evaluation gives
\begin{eqnarray}
&& \langle J_{\bf k} \left( \tau' \right) \, J_{\bf k'} \left( \tau'' \right) \rangle = \frac{\alpha^2 \delta^{(3)} \left( {\bf k} + {\bf k'} \right)}{8 f^2 a \left( \tau' \right) a \left( \tau'' \right)}
\int \frac{d^3 q}{\left( 2 \pi \right)^3} \,  \left[ 1 + \frac{\vert {\bf q} \vert^2 - {\bf q} \cdot {\bf k}}{\vert {\bf q} \vert \, \vert {\bf k} - {\bf q} \vert} \right]^2 \nonumber\\
&&\quad\quad\quad\quad\quad\quad\quad\quad\quad\quad\quad\quad\quad\quad\quad\quad \times 
{\cal A} \left[ \tau' ,\, \vert {\bf q} \vert ,\, \vert {\bf q} - {\bf k} \vert \right] \, {\cal A}^* \left[ \tau'' ,\, \vert {\bf q} \vert ,\, \vert {\bf q} - {\bf k} \vert \right] 
\end{eqnarray}
where
\begin{equation}
{\cal A} \left[ \tau' ,\, \vert {\bf q} \vert ,\, \vert {\bf q} - {\bf k} \vert  \right] \equiv \vert {\bf q} \vert  A_+' \left( \tau' , \vert {\bf q} - {\bf   k} \vert \right) \, A_+ \left( \tau' , \vert {\bf q} \vert \right) + \vert {\bf q} - {\bf k} \vert A_+' \left( \tau' , \vert {\bf q} \vert  \right) \, A_+ \left( \tau' ,\, \vert {\bf q} - {\bf k} \vert \right)
\label{calA}
\end{equation}
and where the relations
\begin{equation}
\left| \vec{\epsilon}_+ \left( {\bf q} \right) \cdot \vec{\epsilon}_+ \left( {\bf k} - {\bf q} \right) \right|^2 
  = \frac{1}{4}  \left[ 1 + \frac{\vert {\bf q} \vert^2 - {\bf q} \cdot {\bf k}}{\vert {\bf q} \vert \, \vert {\bf k} - {\bf q} \vert} \right]^2 
\;\;,\;\;
\vec{\epsilon}_\lambda(-{\bf k}) = \vec{\epsilon}_\lambda^\star({\bf k})
\end{equation}
have been used.

Putting all together, we have
\begin{eqnarray}
\langle \zeta_{\bf k} \left( \tau \right) \zeta_{\bf k'} \left( \tau \right) \rangle & \cong  &
\frac{\alpha^2 \, H^6}{4 \, f^2 \dot{\phi}^2} \, \frac{\left( - k \tau \right)^{n_s-1}}{k^3} \, \delta^{(3)} \left( {\bf k} + {\bf k}' \right) \,
\int \frac{d^3 q}{\left( 2 \pi \right)^3} \,  \left[ 1 + \frac{\vert {\bf q} \vert^2 - {\bf q} \cdot {\bf k}}{\vert {\bf q} \vert \, \vert {\bf k} - {\bf q} \vert} \right]^2  \nonumber\\
&&
\quad\quad\quad\quad \times \Big\vert \int_{-\infty}^\tau d \tau' \left( - \tau' \right) \, {\rm Im } \left[ \varphi_k \left( \tau' \right) \right] {\cal A} \left[ \tau' \, \vert {\bf q} \vert ,\, \vert {\bf q} - {\bf k} \vert \right] \Big\vert^2
\label{2point-intermediate}
\end{eqnarray}

Following similar arguments to those made in subsection \ref{subsec:backreaction}, we can use the approximation (\ref{lorenzo}) for the gauge field mode functions (we discuss this step more in details at the end of this subsection). This leads to
\begin{eqnarray}
\langle \zeta_{\bf k} \left( \tau \right) \zeta_{\bf k'} \left( \tau \right) \rangle & \cong  &
\frac{ \alpha^2 H^6 \, {\rm e}^{4 \pi \xi}}{2^8 \, \pi^3 \,  f^2 \dot{\phi}^2} \,  \frac{\left( - k \tau \right)^{n_s-1}}{k^3} \, \delta^{(3)} \left( {\bf k} + {\bf k}' \right) \nonumber\\
&&\!\!\!\!\!\!\!\!\!\!\!\!\!\!\!\!\!\!\!\!\!\!\!\!\!\!\times \,  \int d^3 q_* \, \left[ 1 + \frac{\vert {\bf q_*} \vert^2 - {\bf q_*} \cdot {\hat k}}{\vert {\bf q_*} \vert \, \vert {\hat k} - {\bf q_*} \vert } \right]^2 \, \vert {\bf q_*} \vert^{1/2} \, \vert {\bf q_*} - {\hat k} \vert^{1/2} \left[ \vert {\bf q_*} \vert^{1/2} + \vert {\bf q_*} - {\hat k} \vert^{1/2} \right]^2 \nonumber\\
&& \times  {\cal I}^2 \left[ 2 \, \sqrt{2 \xi} \left( \sqrt{ \vert {\bf q_*} \vert \,  } + \sqrt{ \vert {\bf q_*} - {\hat k} \vert } \right) \right] 
\label{2point-intermediate2}
\end{eqnarray}
where the integration variable ${\bf q_*} \equiv {\bf q} / \vert {\bf k} \vert$ is dimensionless, and where
\begin{equation}
{\cal I} \left[ z \right] \equiv \sqrt{\frac{\pi}{2}} \, \int_{-k \tau}^\infty d x \, x^{3/2} \, {\rm Re } \left[ H_\nu^{(1)} \left( x \right) \right]
\, {\rm e}^{-z \sqrt{x}}
\label{cali}
\end{equation}
(notice that $x \equiv  - k \tau$). As we are interested only in super horizon modes, $- k \tau \ll 1$, we can set to zero the lower extreme of integration of ${\cal I}$. It is then manifest that we can set $\nu = 3/2$ in the argument of the Hankel function, since the slow roll corrections appearing there only modify (in a negligible amount) the amplitude of the correlator, but not its scale dependence. This leads to
\begin{equation}
{\cal I} \left( z \right) \simeq  \int_0^\infty d x \left( {\rm sin} \, x - x \, {\rm cos } \, x \right) {\rm e}^{-z \sqrt{x}}
\label{cali2}
\end{equation}

For future convenience, we rewrite the correlator (\ref{2point-intermediate2}) as
\begin{equation}
\langle  \zeta_{\bf k}^{\rm inv.decay} \left( \tau \right) \,  \zeta_{\bf k'}^{\rm inv.decay} \left( \tau \right)  \rangle \equiv \frac{2\pi^2}{k^3} (-k\tau)^{n_s-1} \mathcal{P}^2 f_2(\xi) e^{4\pi\xi}  \delta^{(3)}({\bf k}+{\bf k'})
\label{2invdec}
\end{equation}
where $\mathcal{P}$ was defined in (\ref{Pcal}) and the dimensionless function $f_2(\xi)$ is
\begin{eqnarray}
 && f_2(\xi) \equiv  \frac{\xi^2}{8 \, \pi} \,  \int d^3 q_* \, \left[ 1 + \frac{\vert {\bf q_*} \vert^2 - {\bf q_*} \cdot {\hat k}}{\vert {\bf q_*} \vert \, \vert {\hat k} - {\bf q_*} \vert } \right]^2 \, \vert {\bf q_*} \vert^{1/2} \, \vert {\bf q_*} - {\hat k} \vert^{1/2} \left[ \vert {\bf q_*} \vert^{1/2} + \vert {\bf q_*} - {\hat k} \vert^{1/2} \right]^2 \nonumber\\
&& \quad\quad\quad\quad\quad\quad \times \, {\cal I}^2 \left[ 2 \, \sqrt{2 \xi} \left( \sqrt{ \vert {\bf q}_* \vert \,  } + \sqrt{ \vert {\bf q_*} - {\hat k} \vert } \right) \right]     
       \label{f2}
\end{eqnarray}

\begin{figure}
\centerline{
\includegraphics[angle=-90,width=0.7\textwidth]{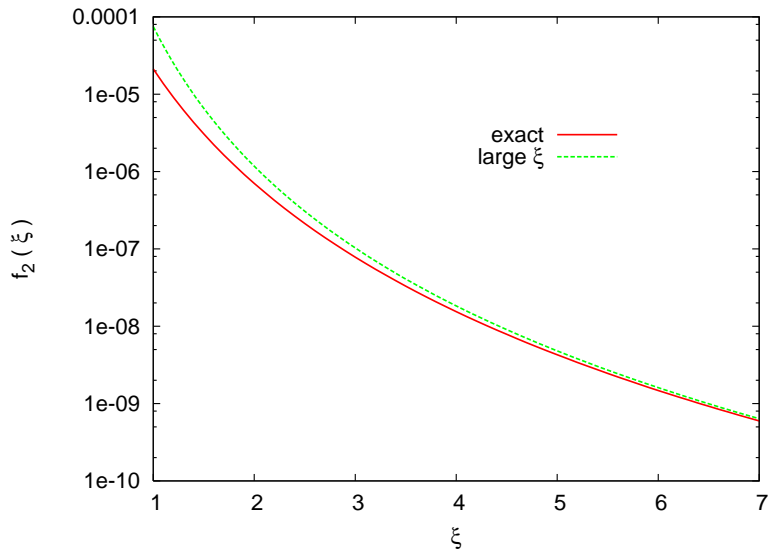}
}
\caption{Exact evaluation and large argument approximation of the function $f_2$.
}
\label{fig:dfdf}
\end{figure}

In general, the function $f_2(\xi)$ needs to be evaluated numerically. However, a simplification is achieved when the argument of ${\cal I}$ is much greater than one. 
\begin{equation}
  {\cal I}\left[ z \right] \cong \int_0^{\infty} dx \frac{x^3}{3} e^{-z\sqrt{x}} = \frac{3360}{z^8} \hspace{5mm}\mbox{for $z \gg 1$}
  \label{I-large}
  \end{equation}
As $\sqrt{ \vert {\bf q}_* \vert \,  } + \sqrt{ \vert {\bf q_*} - {\hat k} \vert } \geq 1$, this approximation is certainly appropriate at large $\xi$. One is then left with a two dimensional integral that numerically evaluates to
\begin{equation}
f_2(\xi) \cong \frac{7.5 \cdot 10^{-5}}{\xi^6} \;\;,\;\; \xi \gg 1
\label{f2-large}
\end{equation}
The degree of accuracy of this approximation can be seen in Figure \ref{fig:dfdf}. It is also useful to have a fit for $f_2$ in the range $2 \leq \xi \leq 3$, as this is the most relevant one for phenomenology (as we discussed in section \ref{sec:pheno}). The best monomial fit to $f_2$ in this range is
\begin{equation}
f_2(\xi) \cong \frac{3 \cdot 10^{-5}}{\xi^{5.4}} \;\;,\;\; 2 \leq \xi \leq 3
\label{f2-fit}
\end{equation}

We conclude this subsection with the justification of the use of  (\ref{mode_soln}) into the integral 
(\ref{2point-intermediate}). We know that the expressions (\ref{mode_soln}) are accurate approximations to the modes $A_+ \left( \tau ,\, k \right)$ in the range $1/8 \xi \leq - \tau \, k \leq 2 \xi$. As discussed 
in subsection  \ref{subsec:backreaction}, for any given mode $k$, this interval corresponds to 
the times for which the amplification of the gauge field is maximal. The key point is to ensure that
there is a common region in the integration space of (\ref{2point-intermediate}) for which both the functions $A_+$ that enter in the expression of ${\cal A}$ can be approximated by (\ref{mode_soln}).
This requires that
\begin{equation}
\frac{1}{8 \xi} \leq - \vert {\bf q} \vert \, \tau'  \leq 2 \xi
\;\;\;\;{\rm and}\;\;\;\;
\frac{1}{8 \xi} \leq  - \vert {\bf q} - {\bf k} \vert \, \tau' \leq 2 \xi
\label{conds}
\end{equation}
or, equivalently,
\begin{equation}
\frac{1}{8 \xi} \leq  \vert {\bf q_*} \vert \, x  \leq 2 \xi
\;\;\;\;{\rm and}\;\;\;\;
\frac{1}{8 \xi} \leq   \vert {\bf q_*} - {\hat k} \vert x \leq 2 \xi
\end{equation}
($x = - k \tau'$ is the integration variable in (\ref{cali}), and ${\bf q_*} = {\bf q} / \vert {\bf k} \vert$). For the super-horizon modes that are relevant for phenomenology, we know that $x$ extends from $0$ to $\infty$. Therefore, for any value of ${\bf q_*}$, there are always values of the rescaled time $x$ for which either $A_+ \left( \tau' ,\, \vert {\bf q_*} \vert \right)$ or $A_+ \left( \tau' ,\, \vert {\bf q_*} - {\hat k} \vert \right)$ is maximal, and approximated by (\ref{lorenzo}). Only when $\vert   {\bf q_*} \vert \simeq \vert {\bf q_*} - {\hat k} \vert$, both conditions  are valid for the same values of $x$, and the the approximations (\ref{mode_soln}) can be used in the whole integrand of (\ref{subsec:backreaction}). Therefore, our result is correct only if the integrand of 
(\ref{subsec:backreaction}) is strongly peaked at $\vert {\bf q_*} \vert = \mathcal{O} \left( 1 \right)$. We have verified with direct inspection that this is indeed the case (We have verified this claim also using the representation (\ref{bessel}) of the gauge field modes which is valid arbitrarily deep in the IR; see Appendix C.)

There is a clear physical reason why the integrand is strongly peaked in the region of phase space where the conditions (\ref{conds}) are satisfied. Notice that, for the values $ z = \mathcal{O} \left( 1 \right) $ which are relevant for the present computation, the expression (\ref{cali2}) has most of its support at $ x = \mathcal{O}\left( 1 \right)$. This means that the ``imprint'' of the fluctuations $\zeta_k$ from the gauge modes occurs when the wavelength of the fluctuation is of the order of the horizon scale ($x$ of order one), and it is caused by the inverse decay of gauge field modes, whose wavelength is also of the order of the horizon scale ($\vert {\bf q_*} \vert $ of order one). This is a  natural outcome of causality/locality. Identical considerations apply in the computation of the three point function that we perform in subsection \ref{sub:3point}.

\subsection{The Primordial Power Spectrum}
\label{sub:power}

The two point correlation function in momentum space is related to the power spectrum by the standard expression
\begin{equation}
  \langle \zeta_{\bf k} \zeta_{\bf k'} \rangle \equiv 
  P_\zeta (k)  \, \frac{2\pi^2}{k^3} \delta^{(3)}({\bf k}+{\bf k'})
\label{power}
\end{equation}

As we have seen in the previous subsection, $\zeta_{\bf k} = \zeta_{\bf k}^{\rm vac} + \zeta_{\bf k}^{\mathrm{inv.decay}}$, and the two terms are uncorrelated. From (\ref{2vacuum}) and (\ref{2invdec}) we get
the power spectrum at late times
\begin{equation}
P_\zeta (k) = {\cal P} \, \left( \frac{k}{k_0} \right)^{n_s-1} \left[ 1 + {\cal P} \, f_2 \left( \xi \right) \, {\rm e}^{4 \pi \xi} \right] 
\label{zz}
\end{equation}
where  $\mathcal{P}^{1/2} = \frac{H^2}{2\pi |\dot{\phi}|}$.

\subsection{Three-Point Correlation Function}
\label{sub:3point}

In this subsection we calculate the three point function of $   \zeta_{\bf k}^{\rm inv.decay} $,
\begin{eqnarray}
  \langle 
\zeta_{\bf k_1} \left( \tau \right) \, 
\zeta_{\bf k_2} \left( \tau \right) \, 
\zeta_{\bf k_3} \left( \tau \right)  
   \rangle
&=& - \frac{H^3}{\dot{\phi}^3} 
\int d \tau_1 \, d \tau_2 \, d\tau_3 \,
\frac{G_{k_1} \left( \tau ,\, \tau_1 \right)}{a(\tau)} \,  \frac{G_{k_2} \left( \tau ,\, \tau_2 \right)}{a(\tau)} \, \frac{G_{k_3} \left( \tau ,\, \tau_3 \right)}{a(\tau)} 
\nonumber \\
&& \quad\quad\quad\quad\quad\quad\quad\quad \times \, \langle J_{\bf k_1} \left( \tau_1 \right) \, J_{\bf k_2} \left( \tau_2 \right)\, J_{\bf k_3} \left( \tau_3 \right) \rangle
\end{eqnarray}
Proceeding as in subsection \ref{sub:2point}, we arrive to
\begin{eqnarray}
\langle 
\zeta_{\bf k_1} \,
\zeta_{\bf k_2} \,
\zeta_{\bf k_3} 
\rangle 
&=& - \frac{\alpha^3 H^9}{f^3 \, \dot{\phi}^3 \, k_1^3 \, k_2^3 \, k_3^3} \,  \delta^{(3)} \left( {\bf k_1} + 
{\bf k_2} +  {\bf k_3} \right) \int \frac{d^3 q}{\left( 2 \pi \right)^{9/2}} \nonumber\\
&&\!\!\!\!\!\!\!\!\!\!\!\!\!\!\!\!\!\!\!\!\!\!\!\!\!\times 
\left[ \vec{\epsilon} \left( {\bf q} \right) \cdot \vec{\epsilon} \left( {\bf k_1} - {\bf q} \right) \right] \,
\left[ \vec{\epsilon} \left( {\bf q} - {\bf k_1} \right) \cdot \vec{\epsilon} \left(  - {\bf q} - {\bf k_3}  \right) \right] \,
\left[ \vec{\epsilon} \left( {\bf q} + {\bf k_3} \right) \cdot \vec{\epsilon} \left(   - {\bf q} \right) \right] 
\nonumber\\
&&\!\!\!\!\!\!\!\!\!\!\!\!\!\!\!\!\!\!\!\!\!\!\!\!\!\times 
 \Pi_i \int_{-\infty}^0 d \tau_i \left[ k_i \tau_i \, \cos \left( k_i \tau_i \right) - \sin \left( k_i \tau_i \right) \right]
\nonumber\\
&&\!\!\!\!\!\!\!\!\!\!\!\!\times {\cal A} \left[ \tau_1 ,\, \vert {\bf q} \vert ,\, \vert {\bf k_1} - {\bf q} \vert \right] \,  
 {\cal A} \left[ \tau_2 ,\, \vert {\bf k_1} - {\bf q}  \vert ,\, \vert {\bf k_3} + {\bf q} \vert \right] \,  
 {\cal A} \left[ \tau_3 ,\, \vert {\bf k_3} + {\bf q}  \vert ,\, \vert  {\bf q} \vert \right] \,  
\label{z3-intermediate}
\end{eqnarray}
where ${\cal A}$ was defined in (\ref{calA}). In this expression we have used the fact that the mode functions $A_+$ are real (which is true both in the approximations (\ref{bessel}) and (\ref{mode_soln})), 
and we have disregarded any (mild) scale dependence.\footnote{Specifically, we have set $\nu=3/2$ in the mode functions entering in the Green function (\ref{green})).}

The correlator depends on the size and the shape of the triangle formed by the vectors ${\bf k_i}$. We denote
\begin{equation}
\vert {\bf k_1} \vert  = k \;\;,\;\;  \vert {\bf k_2} \vert  = x_2 \,  k \;\;,\;\; \vert {\bf k_3} \vert = x_3 \, k
\end{equation}
and, for future convenience, we parametrize the correlator as
\begin{eqnarray}
\langle \zeta_{\bf k_1} \, \zeta_{\bf k_2} \, \zeta_{\bf k_3}  \rangle & \equiv & 
\frac{3  }{80 \left( 2 \pi \right)^{7/2} } \, \frac{\alpha^3 H^9}{f^3 \, \dot{\phi}^3} \, \frac{ {\rm e}^{6 \pi \xi} }{ \xi^3 } \, \frac{\delta^{(3)} \left( {\bf k_1} +  {\bf k_2} +  {\bf k_3}  \right)}{k^6} \, \frac{1+x_2^3+x_3^3}{x_2^3 \, x_3^3} \, f_3 \left( \xi ,\, x_2 ,\, x_3 \right) \nonumber\\
&=& \frac{3}{10} \left( 2 \pi \right)^{5/2} \, {\cal P}^3  \, {\rm e}^{6 \pi \xi} \, 
\frac{\delta^{(3)} \left( {\bf k_1} +  {\bf k_2} +  {\bf k_3}  \right)}{k^6} \, \frac{1+x_2^3+x_3^3}{x_2^3 \, x_3^3} \, f_3 \left( \xi ,\, x_2 ,\, x_3 \right) 
\label{f3-def}
\end{eqnarray}

We proceed from  eq. (\ref{z3-intermediate}) as we did in subsection \ref{sub:2point}. Using the last expression, eq. (\ref{calA}), and eqs. (\ref{lorenzo}), we arrive to
\begin{eqnarray}
f_3 \left( \xi ;\, x_2 ,\, x_3 \right) &=& \frac{5}{3\pi} \, \frac{\xi^3}{x_2 \, x_3 \left[ 1 + x_2^3 + x_3^3 \right]} \, \int d^3 q_* \, \vert {\bf q_*} \vert^{1/2} \, \vert {\hat k}_1 - {\bf q_*} \vert^{1/2} \, 
\vert {\bf q_*} + x_3 \, {\hat k}_3 \vert^{1/2}
\nonumber\\
&& \!\!\!\!\!\!\!\!\!\!\!\!\!\!\!\!\!\!\!\!\!\!\!\! \times \left[ \vec{\epsilon} \left( {\bf q_*} \right) \cdot \vec{\epsilon} \left( {\hat k}_1 - {\bf q_*} \right) \right] \left[  \vert {\bf q_*} \vert^{1/2} + \vert {\hat k}_1 - {\bf q_*} \vert^{1/2} \right] 
{\cal I} \left[ 2 \sqrt{2 \xi}  \left(  \vert {\bf q_*} \vert^{1/2} + \vert {\hat k}_1 - {\bf q_*} \vert^{1/2} \right) \right]
\nonumber\\
&& \!\!\!\!\!\!\!\!\!\!\!\!\!\!\!\!\!\!\!\!\!\!\!\! \times \left[ \vec{\epsilon} \left( {\bf q_*} - {\hat k}_1 \right) \cdot \vec{\epsilon} \left(  - {\bf q_*} - x_3 \, {\hat k}_3  \right) \right] \left[    \vert {\hat k}_1 - {\bf q_*} \vert^{1/2} + 
\vert {\bf q_*} + x_3 \, {\hat k}_3 \vert^{1/2} \right] \nonumber\\
&& \quad\quad\quad\quad\quad\quad\quad\quad\quad\quad\quad\quad
{\cal I} \left[ 2 \sqrt{\frac{2 \xi}{x_2}}  \left(    \vert {\hat k}_1 - {\bf q_*} \vert^{1/2} +  \vert {\bf q_*} + x_3 \, {\hat k}_3 \vert^{1/2} \right) \right] \nonumber\\
&& \!\!\!\!\!\!\!\!\!\!\!\!\!\!\!\!\!\!\!\!\!\!\!\! \times \left[ \vec{\epsilon} \left( {\bf q_*} + x_3 \,  {\hat k}_3 \right) \cdot 
\vec{\epsilon} \left(  - {\bf q_*}   \right) \right] \left[ \vert {\bf q_*} + x_3 \, {\hat k}_3 \vert^{1/2} + \vert  {\bf q_*} \vert^{1/2} \right]  \nonumber\\
&& \quad\quad\quad\quad\quad\quad\quad\quad\quad\quad\quad\quad
{\cal I}  \left[ 2 \sqrt{\frac{2 \xi}{x_3}}   \left( \vert {\bf q_*} + x_3 \, {\hat k}_3 \vert^{1/2} + \vert  {\bf q_*} \vert^{1/2} \right)  \right] 
\label{f3-res}
\end{eqnarray}

To continue, we can set
\begin{eqnarray}
{\hat k}_1 & = & \left( 1 ,\, 0 ,\, 0 \right) \nonumber\\
x_3 {\hat k}_3 & = & - \frac{1}{2} \left( 1 - x_2^2 + x_3^2 ,\, \sqrt{ -  \left( 1 - x_2 + x_3 \right) \left( 1 + x_2 - x_3 \right) \left( 1 - x_2 - x_3 \right) \left( 1 + x_2 + x_3 \right) } ,\, 0  \right) 
\nonumber\\
\end{eqnarray}
and we note that, to a generic vector ${\bf k} = k \left( \sin \theta \, \cos \phi ,\, \sin \theta \sin \phi ,\, \cos \theta \right) $ , corresponds the polarization operator 
\begin{equation}
\vec{\epsilon}_+ \left( {\bf k } \right) =  \frac{1}{\sqrt{2}} \left( \cos \theta \, \cos \phi - i \sin \phi ,\,
\cos \theta \sin \phi + i \, \cos \phi ,\, - \sin \theta \right)
\end{equation}
(it is immediate to verify that this expression satisfies all the properties listed after eq. (\ref{ladder})).

\begin{figure}
\centerline{
\includegraphics[angle=-90,width=0.7\textwidth]{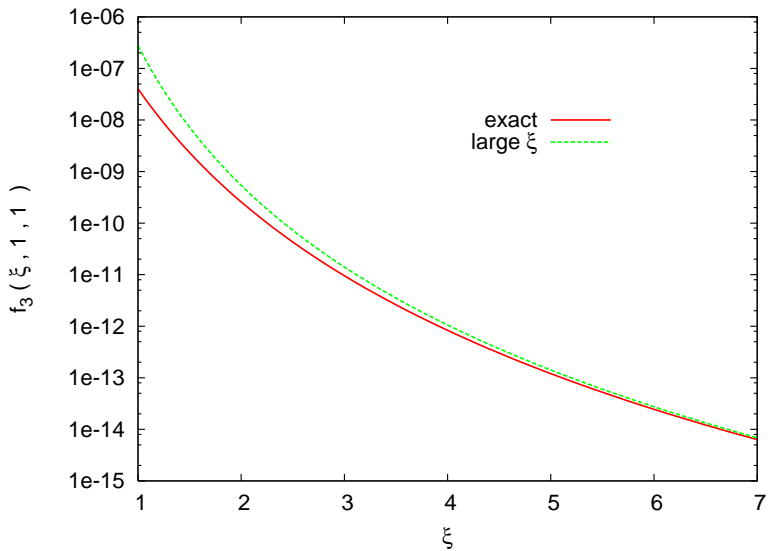}
}
\caption{Exact evaluation and large argument approximation of the function $f_3\left( \xi ;\, 1 ,\, 1 \right)$.
}
\label{fig:dfdfdf}
\end{figure}

In the reminder of this subsection we evaluate $ f_3  $ for the equilateral configuration $x_2 = x_3 = 1$. In the large $\xi$ limit, we can use the analytic approximation  (\ref{I-large}) for ${\cal I}$, and perform the momentum integral numerically. We obtain
\begin{equation}
f_3 \left( \xi ;\, 1 ,\, 1 \right) \cong \frac{2.8 \cdot 10^{-7}}{\xi^9} \;\;\;,\;\;\; \xi \gg 1
\label{f3-large}
\end{equation}
The degree of accuracy of this approximation can be seen in Figure \ref{fig:dfdfdf}. It is also useful to have a fit for $f_3$ in the range $2 \leq \xi \leq 3$, as this is the most relevant one for 
phenomenology (as we discuss in section \ref{sec:pheno}). The best monomial fit to $f_3$ in this range is
\begin{equation}
f_3(\xi ; 1, 1) \cong \frac{7.4 \cdot 10^{-8}}{\xi^{8.1}} \;\;,\;\; 2 \leq \xi \leq 3
\label{f3-fit}
\end{equation}

\subsection{The Bispectrum and Nonlinearity Parameter}
\label{sub:bispectrum}

A popular parametrization of nongaussianity is the nonlinearity parameter $f_{NL}$, introduced by assuming that the curvature perturbation may be expanded as
\begin{equation}
\label{local_first}
\zeta \left( {\bf x} \right) = \zeta_g \left( {\bf x} \right) + \frac{3}{5} \, f_{\rm NL} \,\left[ \zeta_g^2 \left( {\bf x} \right) - \langle  \zeta_g^2 \left( {\bf x} \right) \rangle \right]
\end{equation}
where $\zeta_g(x)$ is a gaussian random field (see \cite{shandera} for a careful discussion of sign conventions).  Both $\zeta$ and $\zeta_g$ may be decomposed as in (\ref{zeta-F}) so that the 
relation between the q-modes of the Fourier decomposition is
\begin{equation}
\zeta_{\bf k} = \zeta_{g,{\bf k}} + \frac{3}{5} \, f_{\rm NL} \, \int \frac{d^3 p}{\left( 2 \pi \right)^{3/2}} \, 
\zeta_{g,{\bf k}} \, \zeta_{g,{\bf k} - {\bf p} } 
\end{equation}
By definition, the three point correlator of $\zeta_g$ vanishes. However, due to the quadratic term in (\ref{local_first}), the three point correlator of $\zeta$ is nonvanishing, and can be 
expressed through a sum of two point correlators of $\zeta_g$.  One finds\footnote{Note that the factors of $(2\pi)^{n/2}$ differ from \cite{small_sound}.  This stems from a different convention
for the normalization of the Fourier transform (\ref{zeta-F}).}
\begin{equation}
\langle \zeta_{\bf k_1} \,  \zeta_{\bf k_2} \,  \zeta_{\bf k_3} \rangle = \frac{ 3 }{ 10 } \, \left( 2 \pi \right)^{5/2} \, f_{\rm NL} P_\zeta \left( k \right)^2 \delta^{(3)} \left( {\bf k_1} + {\bf k_2}  + {\bf k_3} \right) \, \frac{\sum_i k_i^3}{\Pi_i k_i^3}
\label{local_bi_first}
\end{equation}
where the power spectrum was defined in (\ref{power}).
To obtain this expression, recall that the ladder operators are normalized according to (\ref{ladder-phi}), and that the power spectrum $P \left( k \right)$ is related to the two point function as in (\ref{power}). One 
should also  identify the two point function of $\zeta$ with that of $\zeta_g$ (as the difference in subleading in a perturbative expansion), and  disregard the the mild scale dependence of the power spectrum.

By comparison to (\ref{local_bi_first}), one may define an ``effective'' (momentum dependent) nonlinearity parameter, even when the intrinsic nongaussianity is not of the local form (\ref{local_first}).
For axion inflation, using the parametrization (\ref{f3-def}) of the three point correlator, we can write
\begin{equation}
f_{\rm NL}^{\mathrm{eff}}(\xi;x_2,x_3) = \frac{f_3 \left( \xi ;\, x_2 ,\, x_3 \right) \, {\cal P}^3 \, {\rm e}^{6 \pi \xi}}{P_\zeta \left(k \right)^2}
\label{zzz}
\end{equation}
where we recall that ${\cal P}^{1/2} = \frac{H^2}{2\pi|\dot{\phi}|}$.

\subsection{Power spectrum of the Tensor Modes}
\label{sub:GW}

The produced  gauge quanta  also source tensor metric perturbations (gravity waves). The total GW  power spectrum was first given in \cite{ai}. Ref. \cite{Sorbo:2011rz} then pointed out that the chiral nature of the GW produced by the gauge modes can be probed through the resulting nonvanishing $ \langle BE \rangle $ and  $ \langle BT \rangle $ correlators of the CMB, as studied in 
\cite{Saito:2007kt,Gluscevic:2010vv}. For the present model, a positive detection of parity violation would only be possible in a cosmic variance limited experiment, and for a limited portion of the parameter space \cite{Sorbo:2011rz}. In particular, one needs to be in a regime in which the GW production from the gauge modes dominates over that from the vacuum. In the minimal version of the model studied here, this region is ruled out by the nongaussianity limit that we discuss in the next section \cite{ai}. Ref.  \cite{Sorbo:2011rz} circumvented this problem by considering the presence of many (${\cal N} \gsim 10^3$) gauge fields, or by the use of a curvaton.

In the short paper  \cite{ai} we only reported the final result, reserving the present work for the presentation of the details of the computation. As in the meantime this computation has been presented in details in ref.  \cite{Sorbo:2011rz}, we only provide a quick summary here. The tensor modes enter in the spatial components of the  metric as $g_{ij} = a^2 \left( \delta_{ij} + h_{ij} \right)$, with $h_{ii} = \partial_i h_{ij} = 0$. From the Einstein equations, one then finds
\begin{equation}
\frac{1}{2 a^2} \left( \partial_\tau^2  + \frac{2 a'}{a} \, \partial_\tau - \partial_{\bf x}^2 \right) h_{ij} 
    = \frac{1}{M_p^2} \left( - E_i E_j - B_i B_j \right)^{TT}
\label{einsteinh}
\end{equation}
where $TT$ denotes the transverse and traceless projection of the spatial components of the 
energy-momentum tensor  of the gauge field. The computation of the GW production is performed analogously to that of the  density perturbations that we have presented in details in the previous subsections. As for density perturbations, the GW modes produced by the gauge quanta are uncorrelated with those from the vacuum, and the two contributions add up incoherently in the power spectrum. The two GW helicities are obtained from the projectors $\Pi_{ij,L/R} \left( {\bf k} \right) = \epsilon_i^{(\mp)} \left( {\bf k} \right) \,  \epsilon_j^{(\mp)} \left( {\bf k} \right) $. One finds the two power spectra
\begin{equation}
{P}_{L/R} \cong \frac{H^2}{\pi^2 \, M_p^2} \left( \frac{k}{k_0} \right)^{n_T} \left[ 1 + \frac{2 H^2}{M_p^2} \,  f_{h,L/R} \left( \xi \right) \, {\rm e}^{4 \pi \xi} \right]
\label{spehlr}
\end{equation}
where
\begin{eqnarray}
f_{h,L/R} & = & \frac{1}{\xi}   \, \int \frac{d^3 q_*}{\left( 2 \pi \right)^3} \,  \sqrt{  q_*  \, \vert {\hat k} - {\bf q_*} \vert } \;  \frac{ \left( 1 \pm \cos \theta \right)^2 \left( 1 - q_* \cos \theta \pm \sqrt{1 - 2 q_* \cos \theta + q_*^2} \right)^2 }{ 16 \left( 1 - 2 q_* \cos \theta + q_*^2 \right)} \times  \nonumber\\
   &&  \!\!\!\!\!\!  \left\{ \int_0^\infty d x \sqrt{  x }  \left[  \sin x - x \, \cos x  \right] \, 
\left[ \frac{2 \xi}{x} +  \sqrt{ q_*  \, \vert {\hat k} - {\bf q_*}  \vert } \right] 
 {\rm e}^{-2 \sqrt{2 \xi x} \left[ \sqrt{ \vec{q}_* } + \sqrt{\vert {\hat k} - \vec{q}_* \vert} \right]} \right\}^2 
\end{eqnarray}
and where $q_*$ and $\theta$ are, respectively, the magnitude of the  (dimensionless) integration momentum ${\bf q}_*$, and the angle between this vector and the momentum ${\bf k}$ of the mode.

In eq. (\ref{spehlr}), $n_T = - 2 \epsilon$.  We note that the contribution to the spectrum of the modes from the vacuum, and of the modes sourced from the gauge field have the same scale dependence. The reason for this is that the scale dependence of the second term originates from the homogeneous solutions of (\ref{einsteinh}) - which are the vacuum solutions - employed in the Green function. For the same reason, also the two terms in the scalar power spectrum have identical scale dependence,  see subsection \ref{sub:2point}.

\begin{figure}[h!]
\centerline{
\includegraphics[angle=-90,width=0.7\textwidth]{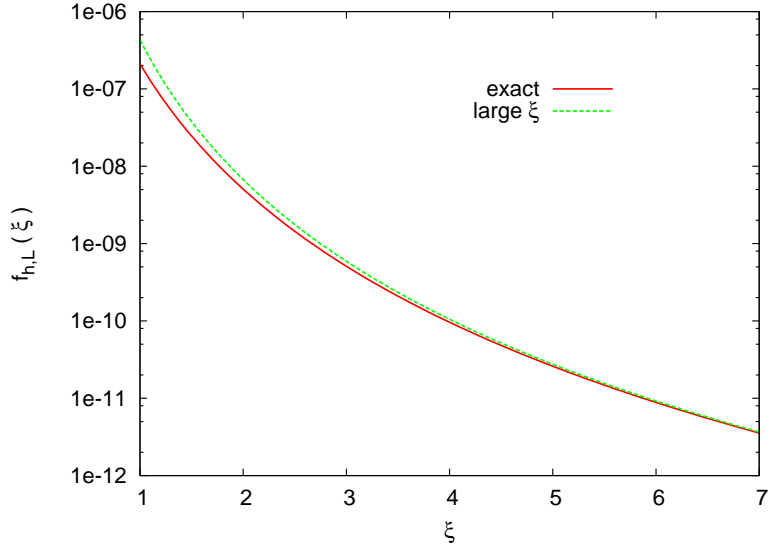}
}
\caption{Exact evaluation and large argument approximation of the function $f_{h,L} \left( \xi  \right)$.
}
\label{fig:fhl}
\end{figure}

At large $\xi$, the integral over $x$ can be performed as in the previous subsections, and one finds  \cite{Sorbo:2011rz} 
\begin{equation}
f_{h,L} \cong \frac{ 4.3 \cdot 10^{-7} }{ \xi^6 } \;\;,\;\;
f_{h,R} \cong \frac{ 9.2 \cdot 10^{-10} }{ \xi^6 } \;\;,\;\;
\xi \gg 1
\label{fhl-large}
\end{equation}
The degree of accuracy of this approximation can be seen in Figure \ref{fig:fhl}. It is also useful to have a fit for $f_{h,L}$ in the range $2 \leq \xi \leq 3$, as this is the most relevant one for phenomenology (as we discussed in section \ref{sec:pheno}). The best monomial fit to $f_{h,L}$ in this range is
\begin{equation}
f_{h,L}(\xi) \cong \frac{2.6 \cdot 10^{-7}}{\xi^{5.7}} \;\;,\;\; 2 \leq \xi \leq 3
\label{fhl-fit}
\end{equation}

From (\ref{fhl-large}) we see that, for the purpose of computing the tensor-to-scalar ratio, one can  disregard the right helicity GW modes 
produced by the gauge fields. One then obtains the result 
\begin{equation}
{P}_{GW} = {P}_{h,L} +   {P}_{h,R} \cong  \frac{2 H^2}{\pi^2 \, M_p^2} \left( \frac{k}{k_0} \right)^{n_T} \left[ 1 +  \frac{H^2}{M_p^2} \,  f_{h,L} \left( \xi \right) \, {\rm e}^{4 \pi \xi} \right]
\end{equation}
first reported in  \cite{ai}.

\section{Phenomenology of Axion Inflation}
\label{sec:pheno}

From the observational perspective, the key quantities which characterize any model of inflation are the spectrum of scalar and tensor perturbations, $P_\zeta$ and $P_{GW}$, along with the bispectrum of scalar 
perturbations $B_\zeta$ that encodes the leading departures from gaussian statistics.  The explicit computations of these quantities was 
presented in Section \ref{sec:correlators}. In this Section, we  study the 
resulting observational signatures.

\subsection{COBE Normalization and Spectral Tilt}
\label{sub:COBE}

In subsections \ref{sub:2point} and \ref{sub:power} we found that two uncorrelated terms contribute to the power spectrum in axion inflation.  These are the usual fluctuations generated from the vacuum, along with 
the modes produced by the inverse decay of the gauge quanta excited by the motion of the inflaton. Taking into account both contributions results in a power spectrum of the form
\begin{equation}
P_\zeta (k) = {\cal P} \, \left( \frac{k}{k_0} \right)^{n_s-1} \left[ 1 + {\cal P} \, f_2 \left( \xi \right) \, {\rm e}^{4 \pi \xi} \right] 
\label{pwr}
\end{equation}
where $n_s = 1 + 2\eta - 6 \epsilon$ is the spectral index, the pivot is $k_0=0.002 \,\mathrm{Mpc}^{-1}$, and we have defined
\begin{equation}
  \mathcal{P}^{1/2} \equiv \frac{H^2}{2\pi |\dot{\phi}|}
\label{calP}
\end{equation}
It is worth noting that both terms  in (\ref{pwr}) have the \emph{same} scale dependence.  Thus, we recover the standard prediction for the 
scalar spectral tilt in single field inflation.  The function $f_2(\xi)$ which appears in (\ref{pwr}) is plotted in Fig.~\ref{fig:dfdf}.  
For $\xi \gg 1$ eq. (\ref{f2-large}) provides an asymptotic expression for 
$f_2$ while, on the other hand, eq. (\ref{f2-fit}) provides a good fit in the $2 \leq \xi \leq 3$ interval (as we  discuss below, this is the 
most  interesting  interval for phenomenology). The COBE normalization $P_\zeta(k) \cong 25\cdot 10^{-10}$ is satisfied along the curve
\begin{equation}
{\cal P}_{\rm COBE} \left( \xi \right) \cong  \frac{e^{-4\pi \xi}}{2 f_2(\xi)}\left[ -1 + \sqrt{1 + 10^{-8} f_2(\xi) e^{4\pi\xi}  }  \right]
\label{COBE}
\end{equation}
in the $\xi-{\cal P}$ plane. We see that, at low $\xi$, the contribution to (\ref{pwr}) sourced by the gauge field is subdominant and we 
recover the standard result ${\cal P}_{\rm COBE}^{1/2} \cong 5 \cdot 10^{-5}$.  The two contributions in (\ref{pwr}) become equal at 
$\xi \cong 2.9$.  As $\xi$ is increased, inverse decay effects dominate the spectrum and the value of $\mathcal{P}^{1/2}_{\mathrm{COBE}}$ 
must be (exponentially) decreased to avoid over-producing density perturbations.  

\begin{figure}[h!]
\centerline{
\includegraphics[angle=-90,width=0.7\textwidth]{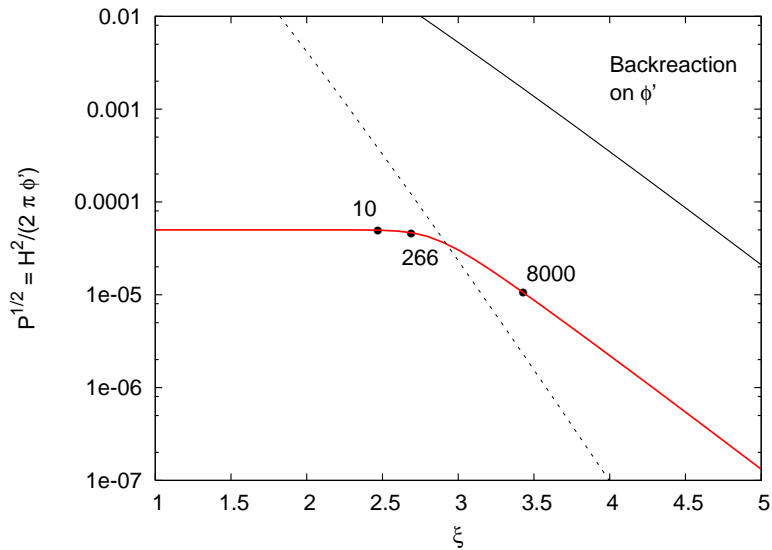}
}
\caption{Values of parameters  leading to the observed COBE normalization of the power spectrum (red line), and  reference  values for the nongaussianity parameter $f_{NL}^{\mathrm{equil}}=10,266,8000$ 
along this curve. See the main text for details.}
\label{fig:cobe}
\end{figure}

The curve (\ref{COBE}) is shown (red solid line) in Figure \ref{fig:cobe}. The black dashed line shown in the Figure separates the (lower) 
region in which the vacuum fluctuations dominate from the (upper) region in which the fluctuations from inverse decay dominate. In the region 
above the solid black line in Fig.~\ref{fig:cobe} the backreaction bound (\ref{back1}) is violated.  In that region of parameter space, the
production of gauge field fluctuations is so strong that dissipative effects (rather than the potential $V(\phi)$) dominate the motion of
$\dot{\phi}$.  From Fig.~\ref{fig:cobe} we see that this backreaction effect can be safely disregarded after the COBE normalization is 
imposed.~\footnote{One should also ensure that the energy density of the produced quanta gives a negligible contribution to the expansion of 
the universe. The resulting condition,  eq. (\ref{back2}), cannot be shown in the  $\xi-{\cal P}$ plane, and therefore needs to be studied on 
a case by case basis. We have verified that this condition is satisfied for the models studied in this section.} 
Finally, Figure \ref{fig:cobe} shows some reference values of $f_{NL}^{\mathrm{equil}}$ along the COBE normalized curve; we discuss this in subsection
\ref{subsec:size}.

\subsection{Tensor-to-Scalar Ratio}
\label{sub:r}

Similarly to the result (\ref{pwr}), the GW spectrum is also the sum of two uncorrelated contributions, one due the modes generated from the 
vacuum, and one due to the modes sources by the gauge field quanta.  In subsection \ref{sub:GW} we derived the result
\begin{equation}
{P}_{GW} \cong  \frac{2 H^2}{\pi^2 \, M_p^2} \left( \frac{k}{k_0} \right)^{n_T} \left[ 1 +  \frac{H^2}{M_p^2}  \,  f_{h,L} \left( \xi \right) \, {\rm e}^{4 \pi \xi} \right]
\label{PGW}
\end{equation}
where $n_T = -2\epsilon$.  The second term in the square braces  corresponds to the gravitational waves sourced by gauge field quanta.  The 
function  $f_{h,L}$ is plotted in Figure \ref{fig:fhl}.  Eq. (\ref{fhl-large}) provides a large argument expansion while 
eq. (\ref{fhl-fit}) provides a fit in the $2 \leq \xi \leq 3$ interval (as we  discuss below, this is the most  interesting  interval 
for phenomenology). 

We define the tensor-to-scalar ratio in the usual way, by normalizing the amplitude of the power in gravitational waves to that in scalar 
fluctuations
\begin{equation}
  r \equiv \frac{P_{GW}}{P_\zeta}
\end{equation}
For $\xi \lsim 1$, inverse decay effects are negligible and we recover the standard consistency relation for $r$, familiar from
single field inflation.  At $\xi \rightarrow \infty$, on the other hand, $r$ tends to a different constant value, which is smaller
than the asyptotic $\xi \rightarrow 0$ value.  From (\ref{pwr}), (\ref{PGW}), and from the result $n_T = -2\epsilon$ it is easy to show that
\begin{equation}
  r = \left\{ \begin{array}{ll}
         -8\, n_T  & \mbox{if $\xi \lsim 3$};\\
        1.8 \, n_T^2 & \mbox{as $\xi \rightarrow \infty$}.\end{array} \right.
\label{consistency}
\end{equation}
Therefore, $r$ is independent of $\xi$ if either vacuum fluctuations or inverse decay effects dominate, while it interpolates between these 
two asymptotic values for intermediate $\xi$.

\subsection{Nongaussianity }
\label{subsec:size}

As discussed in section \ref{sec:overview}, the cosmological fluctuations generated by inverse decay effects are highly nongaussian.
There are many different ways to parametrize departures from gaussianity.  A standard work-horse is the local ansatz:
\begin{equation}
\label{local}
  \zeta(x) = \zeta_g(x) + \frac{3}{5} f_{NL}^{\mathrm{local}} \left[ \zeta_g^2(x) - \langle \zeta_g^2(x) \rangle  \right]
\end{equation}
where $\zeta_g$ is a gaussian random field and $f_{NL}^{\mathrm{local}}$ quantifies the amount of nongaussianity.  Although this simple parametrization has
received considerable attention, it is certainly not the only well-motivated model for a nongaussian curvature perturbation.  More generally, one should consider
the bispectrum, $B_\zeta(k_i)$, which is the 3-point correlation function of $\zeta$ in Fourier space:
\begin{equation}
\label{B}
  \langle \zeta_{\bf k_1} \zeta_{\bf k_2} \zeta_{\bf k_3} \rangle =  B_\zeta(k_i)\, \delta^{(3)}\left({\bf k_1}+{\bf k_2}+{\bf k_3}\right)
\label{def-bisp}
\end{equation}
The bispectrum is a function of three momenta, ${\bf k_i}$, that form a triangle: ${\bf k_1}+{\bf k_2}+{\bf k_3} = 0$.  Hence a generic bispectrum may be characterized by specifying
three interesting properties: the magnitude of the function, its dependence on the shape of the triangle, and its dependence of the size of the triangle.  These properties are usually 
referred to as the \emph{size}, \emph{shape} and \emph{running} of the nongaussianity, respectively.   To characterize these properties, we find it convenient to introduce
\begin{equation}
\vert {\bf k_1} \vert  = k \;\;,\;\;  \vert {\bf k_2} \vert  = x_2 \,  k \;\;,\;\; \vert {\bf k_3} \vert = x_3 \, k
\end{equation}
so that $k$ encodes the overall size of the triangle while the dimensionless quantities $x_2$, $x_3$ encode its shape.

If we assume a local ansatz (\ref{local}) then the bispectrum has a very particular dependence on momenta:
\begin{eqnarray}
  B_\zeta^{\mathrm{local}}(k_i) &=& \frac{3}{10} \left(2\pi\right)^{5/2} f_{NL}^{\mathrm{local}} P_\zeta(k)^2 \frac{\sum_i k_i^3}{\prod_i k_i^3}  \nonumber \\
  &=& \frac{3(2\pi)^{5/2}}{10} \frac{P_\zeta(k)^2}{k^6}\frac{1+x_2^3+x_3^3}{x_2^3x_3^3} f_{NL}^{\mathrm{local}} \label{local_bi}
\end{eqnarray}
This function peaks in the squeezed limit where one of the wave-numbers is much smaller than the other two (for example $k_1 \ll k_2, k_3$). 

\subsubsection{The Size and Running of the Nongaussianity}
\label{subsub:size}

The bispectrum from axion inflation contains two uncorrelated contributions corresponding, respectively, to the usual vacuum fluctuations and to the fluctuations generated by inverse decay processes.  As is well known,
the former contribution gives rise to undetectably small nongaussianity and may be ignored.  The second contribution, however, is more interesting and this was computed in subsection \ref{sub:bispectrum}.
We found that the bispectrum from axion inflation is very different from the local form (\ref{local_bi}).  The bispectrum from axion inflation
peaks on equilateral, rather than squeezed, triangles.  Nevertheless, it is conventional to characterize the size of nongaussianity by matching to (\ref{local_bi})
on equilateral triangles $|{\bf k}_1|=|{\bf k}_2|=|{\bf k}_3|$.  Proceeding in this way we find 
\begin{equation}
f_{\rm NL}^{\mathrm{equil}} \equiv \frac{f_3 \left( \xi ;\, 1 ,\, 1 \right) \, {\cal P}^3 \, {\rm e}^{6 \pi \xi}}{P_\zeta \left(k \right)^2}
\label{eq:fnl-equil}
\end{equation}
The function  $f_3 \left( \xi ;\, 1 ,\, 1 \right)$   is plotted in Figure \ref{fig:dfdfdf}. Eq. (\ref{f3-large}) provides a large argument expansion of this function, while eq. (\ref{f3-fit}) provides a fit in the 
$2 \leq \xi \leq 3$ interval. We stress that this result does not include the negligible contribution from $\langle \delta\phi_{\mathrm{vac}}^3 \rangle$ and is accurate as long as $|f_{NL}^{\mathrm{equil}}| \gsim 1$.

\begin{figure}[h!]
  \centering
  \includegraphics[width=0.5\textwidth,angle=-90]{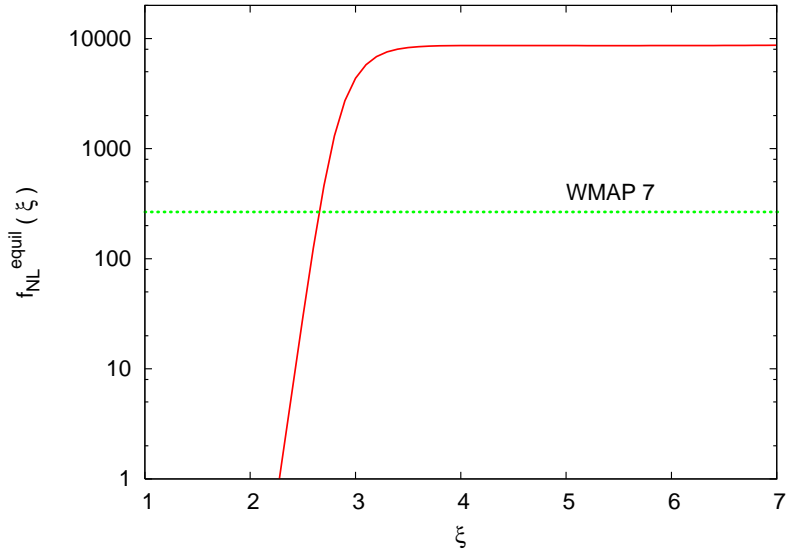}
\caption{Value of $f_{\rm NL}^{\mathrm{equil}} \left( \xi \right)$ due to the perturbations $\zeta$ sourced by the gauge quanta. Also shown is the $95 \%$ CL upper bound from WMAP 7  \cite{wmap7}. 
\label{fig:fnlxi}}
\end{figure}

From eqs.  (\ref{COBE}) and (\ref{eq:fnl-equil}), from the observed power $P_\zeta \cong 25 \cdot 10^{-10}$ (we disregard the slow-roll suppressed scale dependence of $f_{NL}$), and from the expressions
of $f_2$ and $f_3$ computed in the previous section, we can compute the value of $f_{\rm NL}^{\mathrm{equil}} \left( \xi \right)$ along the COBE normalized curve. We show this  in Figure \ref{fig:fnlxi} 
(see also the reference values shown in Figure \ref{fig:cobe}). We notice that $f_{\rm NL}^{\mathrm{equil}} \left( \xi \right) $ saturates to a constant value at large $\xi$ (in the region where the vacuum 
contribution to $P_\zeta$ is negligible). From the large value asymptotic expressions (\ref{f2-large}) and (\ref{f3-large}), we find $f_{\rm NL}^{\mathrm{equil}} \left( \infty \right) \cong 8,600  $. This 
value is already above the $95 \%$ CL upper $-214 <  f_{NL}^{\mathrm{equil}} < 266$ obtained from the WMAP 7 data  \cite{wmap7}. This limit rules out $\xi \gsim 2.65 $. The Planck satellite is expected to be 
able to resolve $f_{NL}^{\mathrm{equil}}$ to $\mathcal{O}(10)$. 

Having quantified the size of nongaussianity in axion inflation, we now turn our attention to its running.
From eq. (\ref{f3-def}) we can see that
\begin{equation}
\label{scaling_bi}
  B_\zeta(k_1,k_2,k_3) = k^{-6} B_\zeta(1,x_2,x_3)
\end{equation}
disregarding the mild, slow roll suppressed scale dependence of the vacuum solutions. The overall $k^{-6}$ behavior reflects
the near scale invariance of the bispectrum from axion inflation.  Slight departures from scale invariance are quantified by the index $n_{NG} -1 = \frac{d \ln |f_{NL}|}{d\ln k}$ which is easily seen to be
 proportional to the slow roll parameters $\epsilon,\eta$, and hence negligible whenever the  observational bounds on $n_s$ are satisfied.  We conclude that the running of nongaussianity is uninterestingly
small in axion inflation.

\subsubsection{The Shape of the Nongaussianity}

In order to discuss the shape of the bispectrum, it is natural to extract the strong $k^{-6}$ scaling 
in (\ref{scaling_bi}) and define a ``shape function'' of the form
\begin{equation}
  S(k_i) = N (k_1k_2k_3)^2 B_\zeta(k_i)
\end{equation}
where the constant of proportionality, $N$, is arbitrary.  This shape function coincides with the quantity that was plotted in many previous works, including \cite{shape} for example.  For the case of interest,
we have
\begin{equation}
S \left( \xi; x_2 ,\, x_3 \right) \equiv  \frac{1+x_2^3+x_3^3}{x_2\,x_3} \, \frac{ f_3 \left( \xi ;\, x_2 ,\, x_3 \right) }{ 3 \, f_3 \left( \xi ;\, 1 ,\, 1 \right) }
\label{eq-shape}
\end{equation}
which is normalized so that $S(1,1)=1$.  Note that the bispectrum is defined only in  the region $x_2 + x_3 \geq 1$, which follows from the triangle inequality.  Moreover, the bispectrum is 
symmetric under interchange of any two momenta, and therefore we can restrict to the region $x_3 \leq x_2 \leq 1$  to avoid considering the same configuration more than once.

\begin{figure}[h!]
\centerline{
\includegraphics[width=0.4\textwidth,angle=-90]{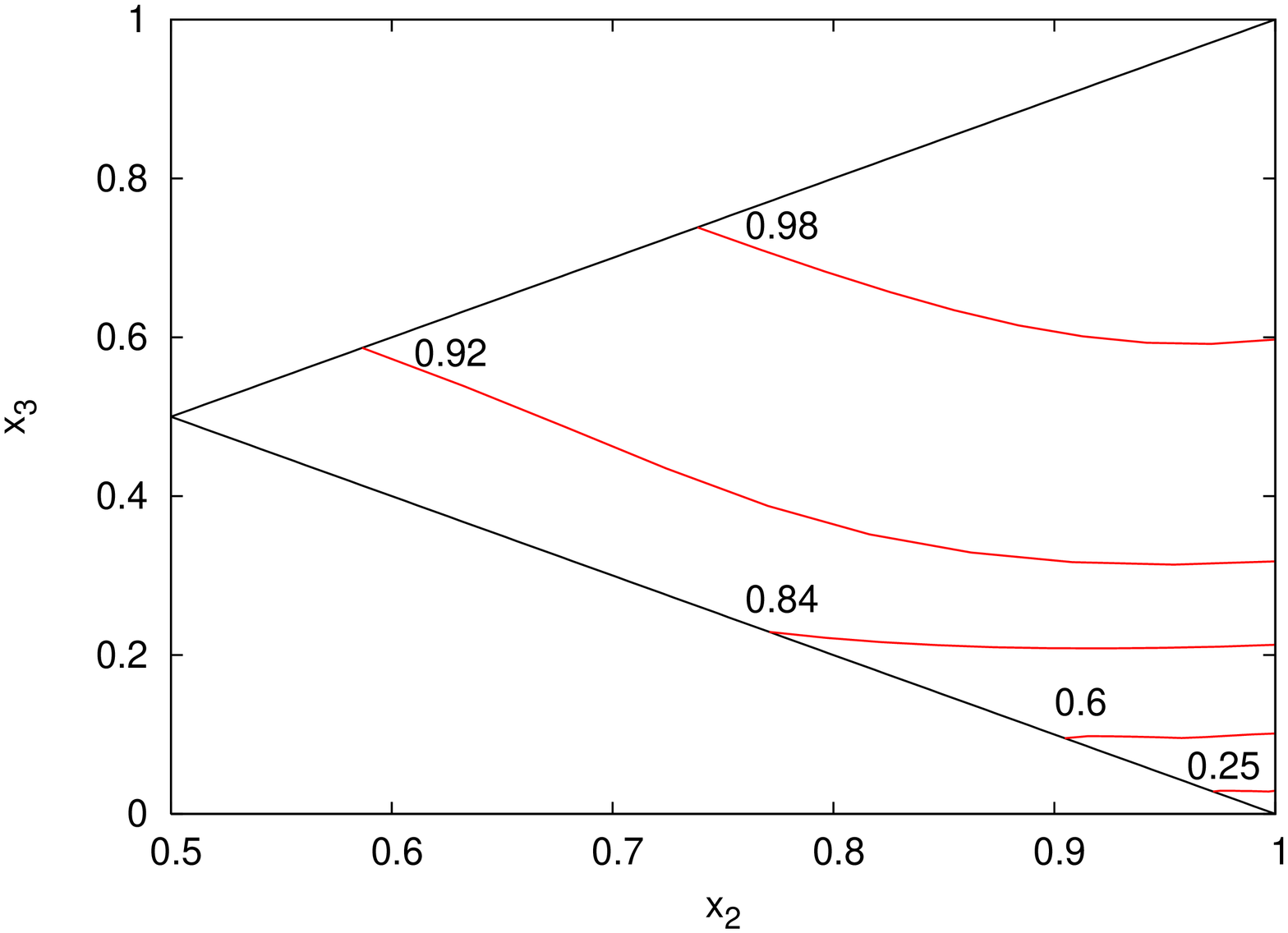}
\includegraphics[width=0.4\textwidth,angle=-90]{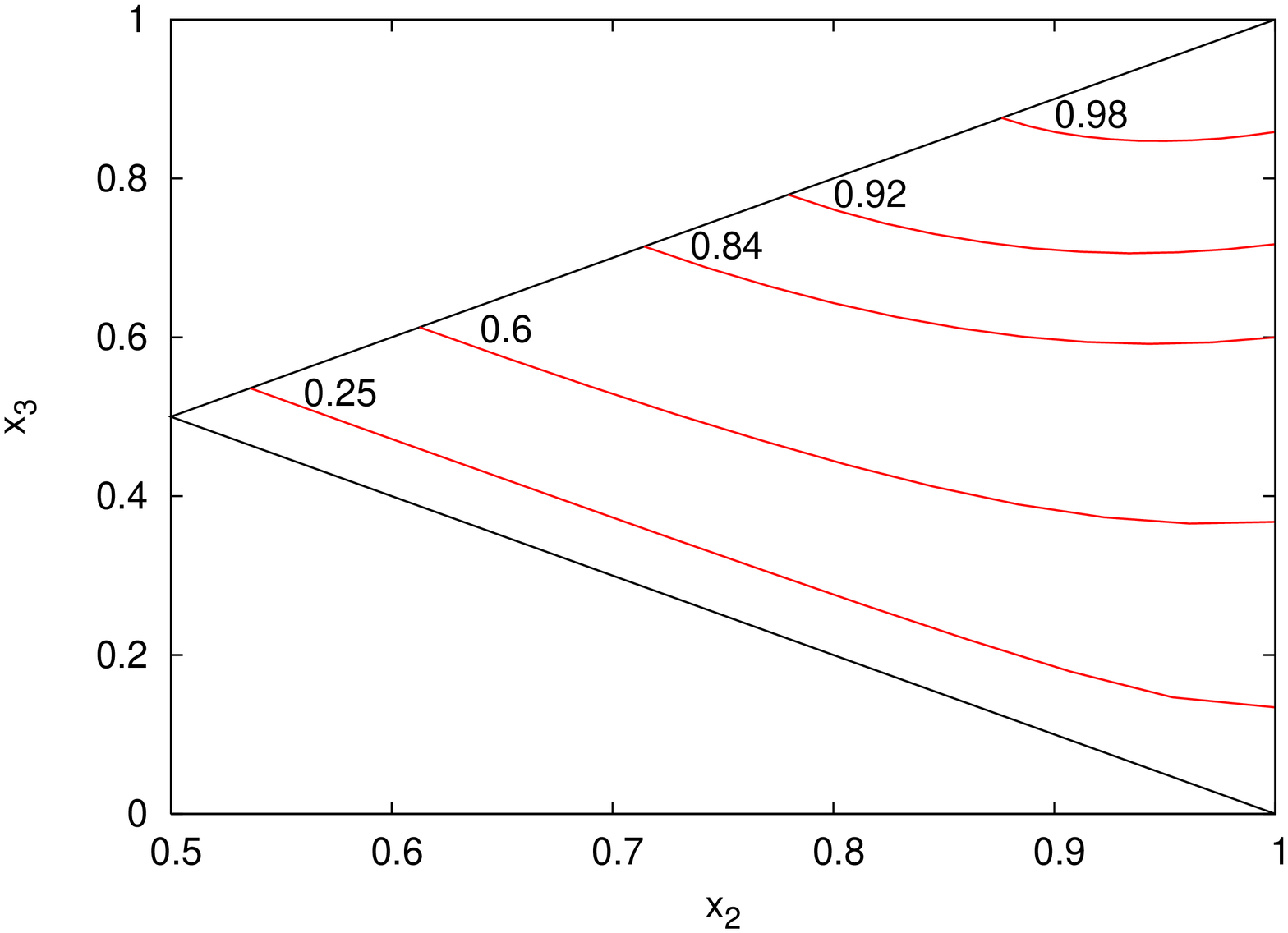}
}
\caption{In the left panel we plot the shape function $S \left( x_2 ,\, x_3 \right)$ in axion inflation, showing that this 
peaks on equilateral triangles. We work in the limit $\xi \rightarrow \infty$, however, this Figure would be nearly indistinguishable had we 
chosen $\xi = \mathcal{O}(1)$.  In the right panel, for comparison, we plot the analogous shape function obtained from the standard equilateral template.  
\label{fig:shape}}
\end{figure}

We plot the shape function $S(x_2,x_3)$ from axion inflation in the left panel of Fig.~\ref{fig:shape}.  The bispectrum in this model depends 
on the parameter $\xi$.  In practice, however, we find that only the size of the nongaussianity (quantified by $f_{NL}^{\mathrm{equil}}$) 
depends strongly on $\xi$.  The shape function $S(x_2,x_3)$, on the other hand, is very mildly dependent on $\xi$.  In Fig.~\ref{fig:shape}
we work in the $\xi \rightarrow \infty$ limit, in which case the shape becomes independent of model parameters.  (This can be seen by using the 
large argument expansion (\ref{I-large}) of ${\cal I}$ in the expression (\ref{f3-res}) for $f_3$.)  For $\xi \sim \mathcal{O}(1)$ this figure would be nearly indistinguishable.

From Fig.~\ref{fig:shape} we see that the bispectrum from axion inflation peaks on equilateral triangles (corresponding to $x_2=x_3=1$) and 
is thus qualitatively similar to the so-called equilateral template which is often employed to analyze CMB data \cite{fnlbounds,wmap7}
\begin{equation}
\label{equil_temp}
  B_{\mathrm{equil}}(k_i) \propto -\frac{1}{k_1^3k_2^3} -\frac{1}{k_1^3k_3^3} -\frac{1}{k_2^3k_3^3} - \frac{2}{k_1^2k_2^2k_3^2} + \frac{1}{k_1k_2^2k_3^3} + (\mathrm{5}\hspace{2mm}\mathrm{perms})
\end{equation}
(where the permutations only act on the last term). Equation (\ref{equil_temp}) is the template that is used to obtain the WMAP7 limit on $f_{\rm NL}^{\mathrm{equil}} < 266$ which we employed in 
subsection \ref{subsub:size}.  The shape function associated with the template (\ref{equil_temp}) is plotted in the right panel of Fig.~\ref{fig:shape} for comparison with the analogous 
result from axion inflation.

To quantitatively compare the bispectrum from axion inflation to the equilateral template (\ref{equil_temp}) we follow \cite{shape} and define a scalar product between any two bi-spectra as 
\begin{equation}
\label{dot}
  B_1 \cdot B_2 \equiv \sum_{\vec{k}_i} B_1(\vec{k}_1,\vec{k}_2,\vec{k}_3) B_2(\vec{k}_1,\vec{k}_2,\vec{k}_3) / \left( \sigma_{k_1}^2 \sigma_{k_2}^2\sigma_{k_3}^2 \right)
\end{equation}
where $\sigma_{k_i}^2$ is the variance of a given mode and the summation runs over all possible triangles. As shown in \cite{shape}, this product is the best estimator for the overlapping of any two distributions: 
if we assume that the real data have the bi-spectrum $B_1$, the template $B_2$ will produce a higher / lower value of nongaussianity according to how large / small the product (\ref{dot}) is \cite{shape}. To be 
quantitative, one defines the cosine of the ``angle'' between the two bi-spectra as
 \cite{shape}
\begin{equation}
\label{cos}
  \cos\left(B_1,B_2\right) \equiv \frac{B_1 \cdot B_2}{ (B_1\cdot B_1)^{1/2} (B_2\cdot B_2)^{1/2}}
\end{equation}
The cosine can be used to quickly estimate how well the limit given in the literature on some given template applies to a  different shape \cite{shape}.  

We have computed the cosine of the ``angle'' between the bispectrum from axion inflation and the equilateral template for several different values of $\xi$.  These results are reported in Table 1.
There we see that the cosine depends only very weakly on $\xi$ and saturates to a value $\cong 0.93$ in the limit $\xi \rightarrow \infty$.  This confirms quantitatively our previous claim that the
shape of nongaussianity is insensitive to $\xi$.

\vspace{2mm}
\begin{center}
  \begin{tabular}{ c | c | c }
   $\xi$ & $\cos\left(B_{\mathrm{inv.decay}},B_{\mathrm{equil}}\right)$ 
    & $\cos\left(B_{\mathrm{inv.decay}},B_{\mathrm{orth}}\right)$  \\
    \hline
    $2$ & $0.94$ & $-0.093$  \\ 
    $3$ & $0.94$ & $-0.12$  \\ 
    $5$ & $0.93$ & $-0.13$  \\ 
    $\infty$ & $0.93$ & $-0.15$   
 \end{tabular}
\end{center}
\noindent{\scriptsize 
 \textit{Table 1}: Cosine of the ``Overlapping angle'', eq. (\ref{cos}), between the nongaussian shape generated by the inverse decay, and the equilateral (column 2) and orthogonal (column 3) templates, for 
different values of $\xi$.}
\vspace{5mm}

Table 1 shows that $\cos\left(B_{\mathrm{inv.decay}},B_{\mathrm{equil}}\right)$ is very close to unity and hence we expect that the WMAP7 
limit $-214 <  f_{NL}^{\mathrm{equil}} < 266$ can be applied also
to axion inflation (at least to first approximation).  This justifies our interpretation of the observational limit on nongaussianity
in subsection \ref{subsub:size}.

Although the bispectrum from axion inflation is very similar to the equilateral template, the two shapes are not identical.  
It may be interesting to characterize the difference between these two shapes -- indeed, this would become pressing in the event 
that Planck, or some other future mission, should detect a non-vanishing bispectrum on equilateral triangles.  From Figure \ref{fig:shape}, we see that the 
bispectrum from inverse decay mostly differs from the equilateral template for $x_2 \simeq x_3 \simeq 1/2$, corresponding to 
``flattened'' triangles, where one side is twice the length of the other two. For such 
triangles, the bispectrum from inverse decay is significantly greater than the equilateral template.  Hence, this provides a natural limit in which the two shapes may be distinguished.  

The fact that axion inflation gives a large nongaussianity on flattened triangles implies a small but nontrivial overlap with the so-called 
``orthogonal template'' that was introduced in \cite{fnlbounds}:
\begin{equation}
\label{orth_temp}
  B_{\mathrm{orth}}(k_i) \propto -\frac{3}{k_1^3k_2^3} -\frac{3}{k_1^3k_3^3} -\frac{3}{k_2^3k_3^3} - \frac{8}{k_1^2k_2^2k_3^2} + \frac{3}{k_1k_2^2k_3^3} + (\mathrm{5}\hspace{2mm}\mathrm{perms})
\end{equation}
(where the permutations only act on the last term). The corresponding shape function evaluates to $+1$ in the equilateral limit, and to $-2$ along the $x_3=1-x_2$ boundary (which includes the flatten 
triangle configurations).  In Table 1 we have computed the cosine of the angle between the bispectrum from axion inflation and the template (\ref{orth_temp}).
We find $\cos\left(B_{\mathrm{inv.decay}},B_{\mathrm{orth}}\right) \cong - 0.15$ (at large $\xi$, see Table 1 for intermediate $\xi$), while $\cos\left(B_{\mathrm{equil}},B_{\mathrm{orth}}\right) \cong  0.21$. 
Therefore, the overlap with the orthogonal template may provide a useful tool to discriminate observationally between the nongaussianity from 
axion inflation and that of the equilateral template.

\subsection{Large Field Inflation}
\label{subsec:large}

Once the COBE normalization is imposed, the key phenomenological predictions of any inflationary model are the spectral index, $n_s$, the tensor-to-scalar ratio $r$, and the 
nonlinearity parameter $f_{NL}$.  As we discussed previously, the spectral index in axion inflation has the standard form 
$n_s = 1 + 2\eta-6\epsilon$ and requires no further discussion.  The remaining observables depend on the coupling $\alpha/f$, the axion
velocity $\dot{\varphi}$ and the Hubble rate $H$.  Out of these three quantities, we have defined the two combinations
\begin{equation}
\xi \equiv \frac{\alpha \dot{\varphi}}{2 f H}
\;\;,\;\;
{\cal P}^{1/2} \equiv \frac{H^2}{2 \pi \vert \dot{\varphi} \vert}
\label{param_combo}
\end{equation}
Both the two and three point correlation functions of $\zeta$ can be written solely in terms of these two combinations. 
Therefore both the power spectrum and the bi-spectrum are a functions of $\xi$ and ${\cal P}$ only. The COBE normalization fixes 
${\cal P}^{1/2}$ in terms of $\xi$, see Figure \ref{fig:cobe}. Therefore, the predicted nonlinearity parameter $f_{NL}$ is function 
of $\xi$ only, see Figure \ref{fig:fnlxi}.

The tensor to scalar ratio, on the other hand, is a function of a different combination of parameters, see eq. (\ref{PGW}). For this reason, 
we cannot present it in a plot as a function of $\xi$ only.  This can only be achieved once the potential for the axion is specified, since 
this provides one additional (slow-roll) relation between the parameters of the model. 

For a specific choice of inflationary potential $V(\phi)$, it is of interest to determine how the combinations (\ref{param_combo}), along
with the tensor-to-scalar ratio, depend on parameters of the underlying theory.  As we discuss in \cite{ai} and also in Section \ref{sec:models}, 
in the models of axion inflation of interest, the axion/inflaton dynamics effectively occurs as in a large field  inflationary model with potential:
\begin{equation}
V = \frac{\lambda_p}{p} \, \varphi^p
\label{largefieldV}
\end{equation}
where $\lambda_p$ has mass dimension $4-p$. Using slow roll approximation, we find
\begin{equation}
H = \sqrt{\frac{\lambda_p}{3 \, p}} \, \frac{\left( 2 \, p \, N \right)^{p/4}}{M_p^{1-p/2}} \;\;,\;\;
\dot{\phi} = \sqrt{\frac{p}{2 N}} \, H \, M_p \;\;,\;\;
\xi = \frac{\alpha \, M_p}{2 f} \, \sqrt{\frac{p}{2 N}}
\end{equation}
where $N$ is the number of e-folds between the moment the CMB scales left the horizon and the end of inflation. 

\begin{figure}[h!]
  \centering
  \includegraphics[width=0.5\textwidth,angle=-90]{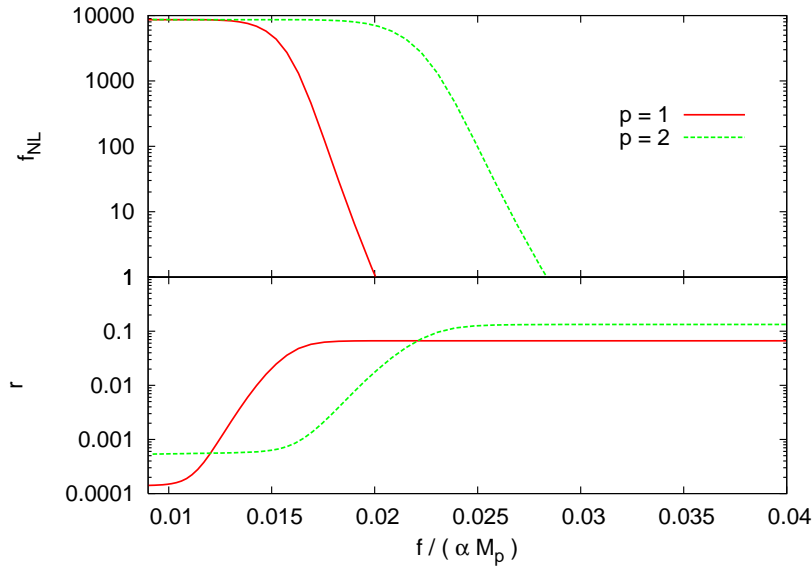}
\caption{Predicted values for the equilateral $f_{NL}$ parameter, and for the tensor to scalar ratio $r$ in axion inflation models, as a function of the 
coupling of the axion to gauge fields, when the axion dynamics is effectively described by a potential $V \propto \varphi^p$, with $p=1,2$.\label{fig:fnl&r}}
\end{figure}

The value of $\lambda_p$ is fixed by COBE normalization and we assume $N=60$ e-foldings of inflation. Then, for any given value of $p$, all 
observational predictions can be written in terms of $f/\alpha$ only. In axion monodromy  \cite{monodromy}, $p=1$; in most of the other 
models one expands the potential close to the minimum, where it is quadratic, $p=2$. We therefore show in Figure \ref{fig:fnl&r} the 
predicted values of $f_{NL}$ and of $r$ for these two cases. We notice that, once $f_{NL}$ is required to be below the WMAP7 bound, the 
standard value for $r$ is recovered.

It is interesting to note that axion inflation models generically predict the \emph{same} values of $n_s$ and $r$ as would be obtained in
vanilla chaotic inflation.  However, our scenario predicts also a large nongaussianity with a (nearly) equilateral shape.  Axion
inflation provides a rare example of a theory which predicts both a detectable tensor-to-scalar ratio \emph{and} a large equilateral 
bispectrum.  Note that, if such a nongaussian signal is eventually detected, then it will immediately fix the value of the coupling 
$\alpha/f$.  On the other hand, if Planck fails to detect nongaussianity then we will have a surprisingly stringent bound on the strongest
axion-type couplings between the inflaton and \emph{any} gauge field.

\section{Cosmological Perturbation Theory}
\label{sec:perts}

In section \ref{sec:overview} we summarized the results of \cite{ai}, providing a brief overview of the growth of gauge quanta and the production of inflaton fluctuations via inverse decay processes, 
in the model (\ref{L}).  In section \ref{sec:correlators} we provided a detailed computation of the relevant correlation functions.  There, and also in \cite{ai}, we neglected scalar metric perturbations.
It is intuitively clear that this should provide a sensible approximation when $f_{NL} \gsim 1$, since the nongaussianity associated with gravitational interactions is tiny as compared to the axion-type
interactions that we study (\ref{int}).  In this section, we reconsider the analysis of section \ref{sec:correlators}, this time consistently accounting for scalar metric perturbations.  As would be expected,
there will be no significant change in our key results.

\subsection{ADM Formalism}
\label{subsec:ADM}

We consider the theory (\ref{L}), minimally coupled to Einstein gravity
\begin{equation}
\label{S}
  S = \int d^4 x \sqrt{-g} \left[ \frac{M_p^2}{2}R - \frac{1}{2}\partial_\mu \varphi \partial^\mu \varphi - V(\varphi) - \frac{1}{4} F^{\mu\nu}F_{\mu\nu} 
  - \frac{\alpha}{8f} \varphi \epsilon^{\mu\nu\alpha\beta} F_{\mu\nu} F_{\alpha\beta} \right]
\end{equation}
The Levi-Civita tensor $\epsilon^{\mu\nu\alpha\beta}$ is defined as
\begin{equation}
\label{levi-civita}
  \epsilon^{\mu\nu\alpha\beta} = \frac{1}{\sqrt{-g}} \eta^{\mu\nu\alpha\beta}
\end{equation}
where the alternating symbol $\eta^{\mu\nu\alpha\beta}$ is $+1$ for even permutations of its indices, $-1$ for odd permutations, and zero otherwise.

To study the cosmological perturbations it is convenient to employ the Arnowitt-Deser-Misner (ADM) form of the metric
\begin{equation}
\label{ADM}
  ds^2 = -N^2 dt^2 + h_{ij} (dx^i + N^i dt)(dx^j + N^j dt)
\end{equation}
This parametrization has the advantage that the lapse function, $N$, and shift vector, $N^i$, appear in the action as Lagrange multipliers and
hence are not dynamical degrees of freedom.  We consider only scalar perturbations of the metric, and we employ the following gauge-fixing choices
\begin{eqnarray}
  && \varphi(t,{\bf x}) = \phi (t) + \delta\varphi(t,{\bf x}) \\
  && h_{ij}(t,{\bf x}) = a^2(t)\delta_{ij} \\
  && A_0  = 0
\end{eqnarray}
corresponding to Coulomb gauge for the vector, along with a flat slicing of the space-time (since the linearized solutions are used for $A_\mu$, we can also impose $\partial_i A_i = 0$). 
It is now straightforward to vary the action (\ref{S}) with respect to $N$, $N^i$, yielding the constraint equations.  
These constraints may be solved perturbatively so that $N$, $N^i$ can be eliminated in favor of $\delta\varphi$, $A_i$.
In this manner, we can derive an action which describes the dynamics of the scalar and gauge field fluctuations, and their
leading interactions.  This procedure has been described many times in the literature, following \cite{maldacena}, and we do not report the explicit steps of the computation in full 
detail in this work.  We also note that  the cosmological perturbations in a model very similar to ours (differing only by
the inclusion of the term $f \left( \varphi \right) F^2$ rather than $\varphi F \tilde{F}$) have already been studied in detail in \cite{seery}.

\subsection{Quadratic Action}
\label{subsec:quad}

The first non-trivial contribution in the perturbative expansion of (\ref{S}) is the quadratic action for the fluctuations $\delta\varphi$, $A_i$,
which give rise to the linearized equations of motion.  To leading order in slow roll parameters, we find the following result for the quadratic
terms involving the scalar field fluctuation
\begin{equation}
\label{S2phi}
  S_2^\varphi = \frac{1}{2}\int d\tau d^3x a^2 \left[ (\partial_\tau \delta \varphi)^2  - \partial_i\delta\varphi\partial_i\delta\varphi 
  + \left( a^2 m^2 - 3\frac{\phi^{'2}}{M_p^2} \right) \delta\varphi ^2 \right]
\end{equation}
The dynamics of the gauge field, on the other hand, arise from the quadratic terms in the expansion of the Maxwell term in the action (\ref{S}).  We find
\begin{equation}
\label{S2A}
  S_2^A = \frac{1}{2} \int d\tau d^3 x \left[ A_i' A_i' - \partial_j A_i \partial_j A_i  + \frac{\alpha \phi'}{f} \epsilon_{ijk}A_i \partial_j A_k \right]
\end{equation}

\subsection{Cubic Action}
\label{subsec:cubic}

The leading nongaussian effects are encoded by the cubic terms in the expansion of (\ref{S}).  These may be divided into two categories: interactions
involving 3 scalars, and those involving one scalar and 2 vectors.  In the first category, we find
\begin{equation}
\label{S3phi}
  S_3^{\varphi\varphi\varphi} = \int d\tau d^3x  \frac{a^2 \phi'}{4\sH M_p^2} \left[ 2 \delta\varphi' \partial_i\partial^{-2}(\delta\varphi') \partial_i(\delta\varphi)
 - \delta\varphi(\delta\varphi')^2 -\delta\varphi \partial_i(\delta\varphi)\partial_i(\delta\varphi)   \right]
\end{equation}
where we have introduced the inverse Laplacian operator, defined through the relation  $\partial^{-2} \partial_i\partial_i f = f$.  
As expected, this coincides with the well-known result for single field inflation \cite{seerymulti,differentKG}.
In (\ref{S3phi}) we have neglected a term proportional to $V'''$, which is usually sub-dominant in the slow roll expansion (see \cite{KGNG} for more discussion).

The interactions (\ref{S3phi}) would be present even in the absence of the gauge field $A_\mu$.  The contribution to the non-linearity parameter due to these terms
is proportional to slow-roll parameters \cite{maldacena,seerylidsey,KGNG} and may therefore be neglected whenever $f_{NL} \gsim \mathcal{O}(1)$.  In this work, we
are therefore justified to ignore (\ref{S3phi}).

Much more interesting for our analysis is the second category of interaction: those involving one scalar $\delta\varphi$ and two gauge fields $A_i$.  From the expansion 
of the Maxwell term $F^2$ in the action (\ref{S}) we find
\begin{equation}
\label{S3max}
  S_3^{\varphi A A} \supset \int d\tau d^3 x \,\frac{\phi'}{ \sH M_p^2}\, \delta\varphi \,\left[ -\frac{1}{4}A_i'A_i' - \frac{1}{8}F_{ij}F_{ij} 
+ \frac{1}{2} \partial^{-2}\partial_\tau\partial_i (F_{ij}A_j')  \right]
\end{equation}
which agrees with \cite{seery}.  Finally, from the last term in (\ref{S}) -- the pseudo-scalar interaction -- we have a contribution
\begin{equation}
\label{S3ax}
  S_3^{\varphi A A} \supset \int d\tau d^3 x \, \frac{\alpha}{f} \,\delta\varphi \,\left[  - \epsilon_{ijk} A_i'\partial_j A_k  \right]
\end{equation}
This interaction gives rise to the source term in (\ref{phi_eqn}) and was accounted for by the analysis of \cite{ai}.
Equations (\ref{S3phi},\ref{S3max},\ref{S3ax}) exhaust all interactions at cubic order.

The interaction (\ref{S3ax}) would be present even in the absence of metric fluctuations and was already accounted for in the analysis of \cite{ai} and section \ref{sec:correlators}; 
this term simply arises from (\ref{int}).  On the other hand, the interactions (\ref{S3phi}) and (\ref{S3max}) are ``new'' and arise due to the consistent inclusion of metric perturbations.  
It is easy to see already that these ``new'' interactions are completely negligible as compared to the pseudo-scalar coupling (\ref{S3ax}).  To see this, note that the strength
of the interaction (\ref{S3ax}) is controlled by the axion decay constant which is
\begin{equation}
  \frac{\alpha}{f} \gsim 10^2\, \frac{1}{M_p}
\end{equation}
whenever the nongaussianity is observationally interesting \cite{ai}.  The analogous coupling strength associated with the interaction (\ref{S3max}) is instead
\begin{equation}
  \frac{\phi'}{\sH M_p^2} \sim \sqrt{\epsilon} \, \frac{1}{M_p} \sim 10^{-1} \, \frac{1}{M_p}
\end{equation}
where in the last approximation we have assumed $n_s - 1 \sim \epsilon \sim 10^{-2}$, which is true in most axion inflation models.  A similar line of reasoning can be used to verify
that the trilinear interactions described by (\ref{S3phi}) are slow roll suppressed with respect to $M_p^{-1}$ and are known to yield a negligible contribution to $f_{NL}$ \cite{KGNG},
as compared to (\ref{S3ax}).

This  analysis  indicates that the inclusion of metric perturbations has no significant impact in the interesting regime where $f_{NL} \gsim \mathcal{O}(1)$.
This intuitive idea has been exploited on numerous occasions in the literature and was recently conjectured as a general decoupling principle in \cite{leblond}.

\subsection{The Equations of Motion}
\label{subsec:KG}

In subsections \ref{subsec:quad} and \ref{subsec:cubic} we derived the leading order contributions to the action for the scalar and gauge field fluctuations.  In order to compute
the nongaussianity in the model, there are two main approaches.  We could use the in-in formalism to compute the bispectrum directly from the interactions 
(\ref{S3phi},\ref{S3max},\ref{S3ax}).  This is the approach that was adopted in \cite{maldacena,seerylidsey,small_sound,seery,seerymulti}.  Alternatively, we can use the perturbed action 
to derive the Klein-Gordon equation, and then compute the bispectrum following \cite{KGNG}.  Both approaches give the same answer.  In our case, we find the second approach to be more
convenient, since it will allow us to take maximal advantage of our previous analysis \cite{ai}.

Variation of the quadratic action (\ref{S2A}) gives the linear equation of motion of the gauge field
\begin{equation}
  \left[ \partial_\tau^2 - \grad^2 -\frac{\alpha \phi'}{f} \grad\times  \right] \vec{A}(\tau,{\bf x}) = 0
\end{equation}
This equation coincides exactly with our previous result (\ref{Amode}).  The consistent inclusion of metric perturbations has no impact on the linear fluctuations of the gauge field,
which is a consequence of the fact that $\langle A_i \rangle = 0$.  

Next, we consider the Klein-Gordon equation for the scalar fluctuation $\delta\varphi$.  Variation of (\ref{S2phi}) along with (\ref{S3phi},\ref{S3max},\ref{S3ax}) gives rise to the 
following equation of motion
\begin{eqnarray}
  &&  \left[ \partial_\tau^2 + 2\sH\partial_\tau - \grad^2 + \left( a^2 m^2  - \frac{3 \phi^{'2}}{M_p^2}\right) \right] \delta \varphi = \frac{\alpha}{f} a^2 \vec{E}\cdot \vec{B} \nonumber \\
&& - \frac{a^2 \phi'}{2 M_p^2} \left[ \frac{\vec{E}^2 +\vec{B}^2 }{2} + \frac{1}{a^4}\partial^{-2}\partial_\tau\left(a^4 \grad\cdot\left(\vec{E}\times\vec{B}\right)  \right)  \right]  
+ \cdots \label{phi_metric}
\end{eqnarray}
where the $\cdots$ on the last line denotes terms of order $(\delta\varphi)^2$ which arise due to variation of (\ref{S3phi}).  As discussed in subsection \ref{subsec:cubic}, those terms
contribute negligibly to the non-linearity parameter and can be ignored.  As a check on our results, we have also derived equation (\ref{phi_metric}) using a completely independent analysis: 
by expanding the Klein-Gordon equation to second order in perturbation theory and using the Einstein constraint equations to eliminate the metric fluctuations, following \cite{malik}.  This 
analysis is briefly summarized in Appendix B.

Equation (\ref{phi_metric}) differs from the result (\ref{phi_eqn}), that was derived by neglecting metric perturbations, in two respects.  First off, equation (\ref{phi_metric}) includes
a slow-roll suppressed correction to the effective mass of the inflaton.  The second discrepancy is the additional contribution to the source on the second line of (\ref{phi_metric}), proportional
to $M_p^{-2}$, arising from variation of (\ref{S3max}).  This contribution is parametrically of order $\sim \sqrt{\epsilon} \frac{\vec{E}^2 + \vec{B}^2}{M_p}$ and may be neglected as compared to the term 
$\frac{\alpha}{f} \vec{E}\cdot \vec{B}$, as discussed in subsection (\ref{subsec:cubic}).  The results of this subsection provide rigorous justification for our previous use of equation (\ref{phi_eqn})
to study the cosmological fluctuations in the model (\ref{S}), in the $f_{NL} \gsim 1$ regime.

\subsection{The Curvature Perturbation}
\label{subsec:zeta}

It now remains to construct the curvature fluctuation on uniform density hypersurfaces, $\zeta$.  Beyond linear order, the relationship between 
$\zeta$ and $\delta\varphi$ is simplest to derive using the $\delta N$-formalism \cite{dN}.  Because the gauge field does not contribute to the expansion 
history of the universe, we have the simple result on large scales:
\begin{eqnarray}
  \zeta \equiv \delta N &=& \frac{\partial N}{\partial \varphi} \delta \varphi + \frac{1}{2} \frac{\partial^2 N}{\partial \varphi^2} (\delta\varphi)^2 + \cdots \nonumber \\
  &=& -\frac{H}{\dot{\phi}} \delta \varphi \left[  1 + \left( 2\epsilon - \eta \right)\frac{1}{2} \frac{H}{\dot{\phi}}\delta\varphi + \cdots  \right] \nonumber \\
 &\cong& -\frac{H}{\dot{\phi}} \delta \varphi \label{zeta_dN}
\end{eqnarray}
where on the last line we have neglected a slow-roll suppressed correction which contributes negligibly to $f_{NL}$ in the case of interest.  

Equations (\ref{phi_metric}) and (\ref{zeta_dN}) are the main results of this section.  Taken together, they provide a rigorous justification for the analysis of \cite{ai}, and of the previous sections,  
which disregard the effect of metric perturbations.

\section{Models of Axion Inflation}
\label{sec:models}

Our results concerning nongaussianity and inverse decay effects are quite general and may be applied directly to a variety axion inflationary 
models.  Moreover, we expect that our qualitative results may have also implications more broadly, for example in any multi-field scenario that involves a 
dynamical axion.\footnote{This seems to be a generic expectation for closed string inflation models.  In the effective SUGRA description the Kahler 
moduli $\tau_i$ are typically paired with axions $\theta_i$ into complex fields $T_i =\tau_i + i\theta_i$.  Absent tuning, one expects the curvature of 
the scalar potential to be comparable in the $\tau_i$ and $\theta_i$ directions.  Hence, nontrivial modular dynamics is typically also associated with 
nontrivial dynamics in the axion sector.  See, for example, reference \cite{kallosh} for a review.\label{foot}}  In this section we survey some interesting microscopic 
constructions, both from field theory and string theory, and comment on the possibility of large nongaussianity.  The key input which our analysis 
requires from a microscopic computation are the inflaton potential $V(\varphi)$, the decay constant $f$, and the dimensionless parameter $\alpha$.  
Together, these determine the quantity
\begin{equation}
  \xi \cong \frac{\alpha M_p^2}{2 f}\left| \frac{V'}{V} \right|
\end{equation}
that measures the strength of inverse decay effects.  As we have seen, inverse decay processes cannot be neglected when $\xi \gsim \mathcal{O}(1)$
which roughly translated into $f / (\alpha M_p) \lsim 10^{-2}$ for the most interesting models; see Figure \ref{fig:fnl&r}. Whenever this inequality is 
satisfied, our findings strongly affect the phenomenology of these models.

In most of the scenarios that we survey, both $V(\varphi)$ and $f$ tend to be fairly well understood while, on the other hand, the coefficient $\alpha$ is rather 
more model dependent (although calculable in principle).  A detailed computation of $\alpha$ in each interesting scenario is beyond the scope of this 
paper and we leave such an analysis to future work.  Although we generically expect $\alpha=\mathcal{O}(1)$,  it should nevertheless be noted that this parameter contributes a 
source of theoretical uncertainty to what follows.  We believe that our phenomenological results should provide a motivation for a detailed microscopic 
computation of $\alpha$ in the various scenarios discussed below.

\subsection{Natural Inflation}

The original natural inflation proposal \cite{natural} was based on the potential\footnote{See \cite{bond} for some proposed particle physics realizations.  
In \cite{kallosh} this model was realized in supergravity -- the so-called ``axion valley'' model.}  
\begin{equation}
  V(\varphi) \cong \Lambda^4\left[ 1 - \cos\left(\frac{\varphi}{f}\right)\right]
\end{equation}
For a spectral index $n_s \gsim 0.95$ this model requires $f \gsim \sqrt{8 \pi} \,  M_p$ \cite{natural2}, for which  the inflaton dynamics shows little difference 
from that standard chaotic inflation with $V(\varphi) \cong \frac{1}{2}m^2 \varphi^2$, at least for the present considerations.  Such 
large values of $f$ weaken the pseudo-scalar coupling $\varphi F\tilde{F}$ and inverse decay effects are negligible unless $\alpha \gsim {\rm few} \times 10^2$.  While 
specific situations have been constructed that can result in a large $\alpha$ - for instance in the some extra-dimensional model  \cite{lorenzo} - 
such  large values conflicts with our general  expectation that $\alpha=\mathcal{O}(1)$.  We conclude that large nongaussianity seems unlikely in the simplest models of natural inflation.

In spite of apparent simplicity, however, the original natural inflation model \cite{natural} seems incompatible with UV completion.  If we interpret $\varphi$ as a 
PNGB then $f > M_p$ suggests a global symmetry broken \emph{above} the quantum gravity scale, where effective field theory is presumably not valid.  
Moreover, $f > M_p$ does not seem possible in a controlled limit of string theory \cite{big_f} (which is the only known  framework wherein such questions 
may be addressed).  Hence, requiring the existence of a sensible UV completion automatically pushes us towards the regime $f \ll M_p$ where inverse 
decay effects are important. We will illustrate how this works in several explicit microscopic realizations below, however, it is clear that this trend applies 
more generally.

\subsection{Double-Axion Inflation}

Perhaps the simplest scenario to realize natural inflation with $f < M_p$ is the double-axion model proposed in \cite{2-flation}.  This model is characterized by 
two axions, $\theta$ and $\rho$, whose potential
\begin{equation}
  V(\theta,\rho) = \sum_{i=1}^2 \Lambda_i^4 \left[ 1 - \cos\left(\frac{\theta}{f_i} + \frac{\rho}{g_i} \right)  \right]
\end{equation}
arises from pseudo-scalar couplings to two different gauge groups: $\frac{\theta}{f_i} F_i \tilde{F}_i$ and $\frac{\rho}{g_i} F_i \tilde{F}_i$.  For 
$f_1g_2=g_1f_2$ one linear combination of $\theta$ and $\rho$ becomes a flat direction of the potential, corresponding to an enhanced symmetry of the 
theory.  Taking $f_1g_2 \cong g_1f_2$ the curvature in this direction becomes controllably flat and one may realize natural inflation with $f_i,g_i \ll M_p$.
This scenario therefore provides a simple and compelling illustration of how the requirement of a UV completion leads to large nongaussianity in natural
inflation.

\subsection{N-flation}

The N-flation model \cite{N-flation} is based on N axion fields $\varphi_i$, each with its own softly broken shift symmetry, resulting in a separable potential
of the form
\begin{equation}
  V(\varphi_i) \cong \sum_i \Lambda_i^4\left[1 - \cos\left(\frac{\varphi_i}{f_i}\right)\right] \cong \sum_i \frac{1}{2} m_i^2 \varphi_i^2
\end{equation}  
where, in the second equality, we have expanded in small field values $\varphi_i \ll f_i$.  For $N \gg 1$ the assisted inflation mechanism \cite{assisted} 
allows inflation to proceed even while all the decay constants are sub-Planckian, $f_i < M_p$.  To a first approximation, the dynamics may be captured by 
a simple quadratic potential $V_{\mathrm{eff}} \cong \frac{1}{2}m^2 \Phi^2$ for the collective field $\Phi^2 \equiv \sum_i \varphi_i^2$ \cite{N-flation2}.    
Successful inflation requires that the collective field traverses a super-Planckian distance in field space, $\Delta\Phi \gsim M_p$.  This is achievable
with sub-Planckian $\varphi_i$ provided the number of axions is sufficiently large.  Roughly speaking we require \cite{kallosh2}
\begin{equation}
  N \sim 240 \left( \frac{M_p}{f_{\mathrm{avg}}} \right)^2
\end{equation}
For $f_{\mathrm{avg}} \sim 10^{-1} M_p$ we have $N \sim 10^3$ while $f_{\mathrm{avg}} \sim 10^{-2} M_p$ would require $N \sim 10^6$.

N-flation makes sense as a purely field theoretical construction, however, much of the interest in this scenario arises because the exponentially
large values of $N$ that are required may be rather generic in string theory compactifications.  In this context, axions arise on dimensional reduction,
from integrating $p$-form gauge potentials over $p$-cycles (see \cite{kallosh,bm} for a review).  For example, in type IIB string theory one has axions
$b_i$ and $c_i$ which arise, respectively, from integrating the Neveu-Schwarz (NS) and Ramond-Ramond (RR) 2-forms $B_{MN}$ and $C_{MN}$ over 
compact 2-cycles.  Generically, instanton effects break the shift symmetry down to a subgroup $b_i \rightarrow b_i + (2\pi) f_i$, leading to periodic 
contributions to the effective potential (similarly for $c_i$).

The general framework described above shows how N-flation may arise within string theory: the low energy theory contains one axion for each 
independent cycle that $B_{MN}$ can wrap and generic Calabi-Yau compactifications may contain exponentially large numbers of such cycles.  There exist know examples with 
$N$ as large as $\sim10^5$ \cite{N105}, however, to our knowledge there is no general theorem that prohibits finding Calabi-Yau manifolds with
even larger values.  

There may be many ways to realize N-flation in an explicit, stabilized string theory compactification.  The first efforts in this direction were undertaken in
\cite{N-flation2} and further analyzed in \cite{kallosh2}.  See also \cite{grimm} for an alternative construction.

In order to quantify inverse decay effects, we are most interested in the effective coupling between the collective field $\Phi$ and a given gauge field.
The detailed derivation of such interactions is rather complicated and dependent on model building details (see \cite{green,multi,Nmag} for a more 
detailed discussion).  We will not attempt to estimate the effective value of $\alpha / f$, but rather note that a generic spectrum is expected to contain 
some axions with $f_i \ll M_p$, hence strong inverse decay effects (and their associated nongaussianities) are at least plausible in N-flation.

\subsection{Axion Monodromy Inflation}

The axion monodromy model \cite{monodromy} is a string theoretic construction based on a single axion field.  The key ingredient in this construction
is a suitably wrapped brane which leads to a non-periodic contribution to the inflaton potential, explicitly breaking the shift symmetry.  This monodromy
in the moduli space allows the axion to develop a kinematically unbounded field range and accommodate the super-Planckian excursions required for
large-field inflation.  In the effective field  theory description of this scenario, the inflaton potential has the form
\begin{equation}
\label{modpot}
  V(\varphi) = \mu^3 \varphi + \Lambda^4 \cos\left(\frac{\varphi}{f}\right)
\end{equation}
where the linear term arises from wrapped branes while the (subdominant) periodic modulation arises from instanton effects.

In \cite{monodromy2} the decay constant for axion monodromy inflation was studied in detail.  It was shown that microphysics bounds the allowed
values as
\begin{equation}
  0.06 \,  \frac{g_s^{1/4}}{\mathcal{V}^{1/2}} < \frac{f}{M_p} < 0.9 \, g_s
\end{equation}
where $g_s < 1$ is the string coupling and $\mathcal{V} \gg 1$ is the compactification volume in string units.  This bound illustrates that $f \ll M_p$ 
in this model.  If $\alpha = \mathcal{O}(1)$ can be realized in a consistent string compactification, then large nongaussianity is easily accommodated 
by axion monodromy inflation.

The small periodic modulation of the potential (\ref{modpot}) is not important for determining the number of $e$-foldings of inflation, 
however, this term may nevertheless have an important role of the phenomenology of the model.  In \cite{monodromy2}, and also in \cite{monodromyNG},
it was shown that axion monodromy  models can give rise to large resonant-type nongaussianities \cite{chen1,chen2,pajer,leblond}.  Depending on the 
parameters, either resonant effects or inverse decay effects may dominate the bispectrum in the regime $f \ll M_p$.  It would be interesting to study the
combined observational impact of these two effects in future work.

\subsection{Axion/4-Form Mixing}

In \cite{kaloper} a realization of chaotic inflation was proposed which shares many features of the axion monodromy model discussed above.  
Here the axion $\varphi$ mixes with a 4-form field strength through a term like $\varphi \, \epsilon^{\mu\nu\alpha\beta} F_{\mu\nu\alpha\beta}$.  The theory also 
includes charged membranes which source a background for the conjugate momenta of the gauge field and break the axion shift symmetry, giving rise to 
a non-periodic potential $V(\varphi) \cong \frac{1}{2} m^2 \varphi^2$ which is robust against a wide variety of corrections \cite{mixing2}.  There is no obstruction to 
realizing inflation with $f\ll M_p$ in this model, and hence we generically expect that inverse decay processes may play an important role.  

As in the case of axion monodromy, the power-law potential will generically be modulated by subdominant oscillatory features arising
from instanton effects.  In \cite{mixing2} nongaussianities were discussed.  In general resonant effects may operate in concert with inverse decay
processes, and it would be interesting to study their combined impact.

\subsection{Dante's Inferno}

In \cite{dante} a model was proposed which consists of two axions, $r$ and $\theta$, with decay constants $f_r < f_\theta \ll M_p$.  It is assumed that a 
linear combination of these receives a periodic potential from nonperturbative effects and, moreover, that some explicit shift symmetry breaking effect
generates a non-periodic contribution $W(r)$.  The effective potential therefore takes the form
\begin{equation}
\label{dantepot}
  V(r,\theta) = W(r) + \Lambda^4\left[1 - \cos\left(\frac{r}{f_r} - \frac{\theta}{f_\theta}\right)\right]
\end{equation}
For a power-law $W(r) = \mu^{4-p}\, r^p$ the dynamics of this model are well approximated by a single field with effective potential 
$V_{\mathrm{eff}} = (f_r / f_\theta)^p \mu^{4-p} \varphi^p_{\mathrm{eff}}$ and our analysis of inverse decay processes is directly applicable.  The possibility of large 
nongaussianity follows immediately from $f_r < f_\theta \ll M_p$.

In \cite{dante} it was shown how to embed the model (\ref{dantepot}) within string theory as a modest extension of axion monodromy
\cite{monodromy,monodromy2}.  It may be possible to realize such a potential from axion/4-form mixing, along the lines of \cite{kaloper,mixing2}.

\subsection{Multi-Field Scenarios}


So far we have focused our attention on models which can, at least to first approximation, be described in terms of the dynamics of a single field
(that may represent a collective excitation or be related non-trivially to the axions of the original theory).  Such scenarios are appealing,
because our analysis of inverse decay processes applies more-or-less without modification.  However, there are also a number of interesting inflation
models which involve dynamical axions but are \emph{not} well described in terms of a single dynamical field.  (As discussed in footnote \ref{foot}, this
scenario seems especially natural in supergravity models.)  In multi-field models with $f \ll M_p$ strong inverse decay effects are clearly possible, 
however, the detailed phenomenology is more complicated than what we have presented.  Nevertheless, we expect that large nongaussianity should
be possible.  It would be interesting to explore multi-field inverse decay processes in future works.

There are a variety of interesting multi-field models involving axions, for example racetrack inflation \cite{racetrack1,racetrack2} and the axionic D3/D7
model \cite{D3D7}.  Possibly,   interesting models from the perspective of obtaining strong inverse decay effects, are based on the large volume
compactification of \cite{LVC1,LVC2,LVC3}.  For example, roulette inflation \cite{roulette} -- which generalizes \cite{CQ} to incorporate the dynamics of the
axionic partner of the Kahler modulus -- is characterized by significant motion in the axion direction during the early stages of inflation.  
See \cite{astro,modular} for a discussion of axion/moduli couplings and decays in large volume inflationary scenarios.  See also \cite{misra} for a related model.

\section{Conclusions}
\label{sec:conclusions}

Probably the greatest difficulty in inflationary model-building is to protect the required flatness of the potential from radiative corrections. One of the simplest 
solutions to this problem is to assume that the inflaton $\varphi$ is a Pseudo-Nambu-Goldstone-Boson (PNGB).  In this case the inflaton enjoys a shift symmetry 
$\varphi\rightarrow \varphi + \mathrm{const}$, which is broken either explicitly or by quantum effects.  In the limit of exact symmetry, the $\varphi$ direction is 
flat and thus loop corrections to the potential are controlled by the smallness of the symmetry breaking. PNGBs, like the axion, are ubiquitous in particle physics: 
they arise whenever an approximate global symmetry is spontaneously broken and are plentiful in string compactifications. The idea of using a PNGB as inflaton was 
first put forward  in \cite{natural}; in this minimal realization a single axion is present; it turns out that this model can produce a sufficiently flat spectrum 
of perturbations only if  the axion decay constant $f$ is above the Planck scale. This regime may be impossible to obtain in a controlled way: one can expect that 
gravity breaks the axionic shift symmetry at a smaller scale and it has also been conjectured that $f>M_p$ cannot be realized in string theory. These problems have 
been solved by a number of controlled realizations in which an axion, or a combination of axions, with an axion scale $f$ a few orders of magnitude smaller than $M_p$, 
behaves effectively as a large field inflaton \cite{2-flation,N-flation, N-flation2,monodromy,monodromy2,kaloper,lorenzo}.

Axions are coupled to gauge fields through $\frac{\alpha}{f} \varphi F {\tilde F}$.  More generally, from the perspective of effective field theory, a coupling 
$\varphi F \tilde{F}$ must be included whenever $\varphi$ is pseudo-scalar.  The dimensionless coupling $\alpha$ is a model-dependent quantity, however, from an 
field theoretical point of view, one does not expect it to be $\ll 1$.  
In this work we studied the phenomenological signatures induced by  this coupling, keeping $\alpha/f$ as a free parameter. The crucial point of our work is that, in 
presence of this coupling, the motion of $\varphi$ induces a tachyonic growth for one polarization $\delta A$ of the gauge field. This amplification is most important 
for modes with physical wavelength comparable to the horizon $\lambda \sim 1/H$. These produced gauge quanta inverse decay to produce inflaton fluctuations of a comparable 
wavelength, consistent with the general expectation from causality/locality. These inflaton modes then leave the horizon, and the corresponding density fluctuations, 
$\zeta_{\rm inv.decay}$, become frozen and contribute to the observable cosmological perturbations.  An analogous process leads to production of gravity waves. These scalar 
and tensor perturbations add up incoherently with those generated by the expansion of the universe (the standard ``vacuum'' modes, $\zeta_{\rm vac}$ and $h_{\rm vac}$). 

We find that the amplitude of the perturbations generated by the inverse decay is an exponentially growing function of $\alpha/f$. These modes dominate over the vacuum ones for 
$\alpha / f \gsim 10^{-2} \, M_p^{-1}$ (the precise value depending on the inflaton potential). Due to the exponential sensitivity on the coupling, there is only a small range in 
$\alpha/f$ for which both $\zeta_{\rm inv.decay}$, and $\zeta_{\rm vac}$ are relevant. For smaller values of $\alpha/f$  this new effect is completely negligible, and the standard 
results are recovered. For larger values, drastically new predictions are obtained. In particular, the main characteristic of $\zeta_{\rm inv.decay}$ is that they are highly 
nongaussian: this is due to the fact that two gauge quanta participate in the inverse decay, and that the initial distribution of these quanta is itself gaussian (loosely speaking, 
$\zeta_{\rm inv.decay}$ behaves as the square of a gaussian field, which is obviously not gaussian).  As a consequence of this effect, the nongaussianity parameter $f_{NL}$ also grows 
exponentially with $\alpha/f$ in the region of parameter space for which both $\zeta_{\rm vac}$ and $\zeta_{\rm inv.decay}$ are comparable. When  $\zeta_{\rm vac}$ can be neglected, 
$f_{NL}$ saturates to about $8,600$ in the equilateral configuration,  well beyond the current WMAP limit $f_{NL}^{\rm equil} < 266$. We find that the WMAP limit allows only values 
of  $\alpha/f$ for which  $\zeta_{\rm inv.decay}$ can contribute  $\lsim 10\%$ to the power spectrum (see Figure \ref{fig:cobe}).  In this regime, also the spectrum of gravity waves 
produced by the inverse decay is much smaller than that from vacuum.

We have seen that nongaussianity from axion inflation is greatest for the equilateral configuration.  This is related to causality/locality of the underlying inverse decay production
mechanism.  To understand this, intuitively, recall that a mode $\zeta_k$ is sourced by the inverse decay of gauge perturbations of comparable wavelength.  Consequently, correlation 
\begin{equation}
\langle \zeta_{k_1} \zeta_{k_3} \zeta_{k_3} \rangle \propto \int d^3 q_1 d^3 q_2 d^3 q_3 \langle \delta A_{q_1}  \delta A_{k_1-q_1}  \delta A_{q_2}  \delta A_{k_2-q_2}  \delta A_{q_3}  \delta A_{k_3-q_3} \rangle
\end{equation}
is suppressed between modes of very different scale. We found that the shape of nongaussianity produced by this mechanism is very well reproduced by the equilateral template that has been widely studied 
in the literature.  Quantitatively, the ``cosine'' between the two shapes (a measure of how much the two shapes coincide \cite{shape}) is $\cong 0.93$. Therefore, it is sensible to use the limits on  
$f_{NL}^{\rm equil} $ (obtained with the use of the equilateral template) to probe  the current mechanism. On the other hand, the specific shape that we have computed is nevertheless distinct from the equilateral
template and this fact might be useful to distinguish axion inflation from other models (for instance, the shape we found has a significant overlapping also with the  orthogonal template).  This issue can be more 
thoroughly  explored when / if a nonzero value for  $f_{NL}^{\rm equil} $ will be found in the data.  The nongaussian signature of axion inflation, together with the requirement of standard results for the spectral 
tilt and the tensor-to-scalar ratio tensor of large field inflation, will allow to falsify the mechanism (or, at least, the simplest version 
that we have studied here) in the near future.  This is due to the fact that large field inflationary model provide much larger - 
and detectable - values of $n_s-1$ and of $r$ than many other classes of models.

We stress that the nongaussianity which we have studied is very different from the so called resonant nongaussianity \cite{chen1,chen2}, 
that has been discussed previously \cite{monodromy2,monodromyNG,pajer,leblond} 
in the context of axion monodromy inflation \cite{monodromy}.  Axion monodromy is one of the particle physics realizations of axion inflation; in this model the periodic potential typical of axions is only a 
subdominant term in the inflaton potential, and it provides a nongaussian modulation of the inflaton perturbations.  Depending on model parameters, either this effect or inverse decay processes may dominate
the bispectrum.

It is remarkable that, in the mechanism we have studied, the large nongaussianity is obtained in a rather minimal way (simply by considering a coupling of the pseudo-scalar allowed by the symmetries of the model, 
and therefore expected in an effective field theory context).  Nongaussianity is a measure of the strength of interactions of the inflaton, which are typically constrained to be small by the requirement of a flat 
potential. To obtain observable nongaussianity, previous studies have invoked non-standard field theories (involving small sounds
speed \cite{small_sound} or higher derivatives \cite{NL}), or initial conditions \cite{small_sound,nonBD1,nonBD2,nonBD}, potentials 
with sharp features \cite{chen1,chen2}, dissipative effects \cite{trapped}, fine-tuned inflationary trajectories \cite{turnNG} or post-inflationary effects (such as preheating \cite{preheatNG,preheatNG2}).  The 
mechanism we have studies circumvents the common lore result  in a very novel way: the interaction that gives rise to large nongaussianity, eqn.~(\ref{int}), does not play any role in the background dynamics and 
is thus unconstrained by the requirement of slow roll.  At the same time, the effect of this interaction 
(namely, the amplification of gauge quanta) persists during the entire inflationary phase, leading to a (nearly) 
scale invariant signature (as opposed to sporadic episodes of particle productions \cite{pp1,pp2,pp3} that would lead to a highly localized nongaussianity in momentum space).

Finally, it is also remarkable that current observational limit on nongaussianity \emph{already} place a surprisingly stringent bound on pseudo-scalar interactions of the form (\ref{int}).
As a comparison, the bound on the coupling of the much lighter QCD (or QCD-like) axion to photons is 
$\alpha_{\rm photons} / f \lsim \mathcal{O} \left( 10^{-11}  \right) \, {\rm GeV}^{-1}$, from energy loss in stars \cite{pdg}. 
On the other hand, the mechanism that we have studied provides the bound  
$\alpha / f \lsim \mathcal{O} \left( 10^{-16}  \right) \, {\rm GeV}^{-1}$ on the coupling of the   pseudo-scalar  inflaton  with {\em any } gauge field.  This provides a unique window for constraining 
-- or perhaps probing -- a large class of inflationary models. As we have reviewed in section \ref{sec:models}, there are a variety of interesting multi-field models involving axions, many of which can be 
realized in string theory. While most of these studies provide the value of axion decay constant $f$ and of the potential $V(\varphi)$, 
comparatively less attention has been devoted to compute the dimensionless coupling, $\alpha$. We believe that our findings provide a 
strong motivation to undertake such a study.

\section*{Acknowledgments}

We thank L.~Sorbo for valuable discussions.  This work was 
supported in part by DOE grant DE-FG02-94ER-40823 at UMN. 

\renewcommand{\theequation}{A-\arabic{equation}}
\setcounter{equation}{0}  

\section*{APPENDIX A: Gauge Field Mode Functions}

In this appendix we discuss the solutions of equation (\ref{Amode}), describing the unstable  growth of gauge field fluctuations in the background of the slowly  rolling inflaton (hence, $\xi$ can be taken constant at leading order in the slow roll parameters).  We require that the gauge field is initially in the adiabatic vacuum:  $A_{\pm}(\tau,k) = e^{-ik\tau} / \sqrt{2k}$ for $k\tau \rightarrow -\infty$. The mathematical properties used here can be found in chapter $14$ of \cite{abramowitz}. The solution of (\ref{Amode}) which satisfies this condition may be expressed in terms of Coulomb functions:
\begin{equation}
\label{coulomb}
  A_{+}(\tau,k) = \frac{1}{\sqrt{2k}} \left[ G_0(\xi,-k\tau) + i F_0(\xi,-k\tau)  \right]
\end{equation}
The production of gauge fluctuations is only interesting in the region of phase space $-k\tau \ll 2 \xi$ and when $e^{\pi \xi} \gg 1$ (sere the discussion in subsection \ref{subsec:backreaction}).  In this regime equation (\ref{coulomb})
may be very well approximated in terms of the modified Bessel function of the second kind, $K_\nu(z)$, as
\begin{equation}
\label{bessel}
  A_{+}(\tau,k) \cong \sqrt{\frac{-2\tau}{\pi}} \, e^{\pi \xi} \, K_{1}\left[2 \sqrt{-2\xi k \tau}\right]
\end{equation}
The growth of the modes (\ref{bessel}) saturates deep in the IR: for $-k\tau \rightarrow 0$ we have $A_{+} \rightarrow e^{\pi \xi} / (2 \sqrt{\pi k \xi})$ so that the physical
electric and magnetic field vectors (\ref{electromagnetic}) decay sufficiently far outside the horizon.  
An inspection of the solutions  shows that the interesting  physical effects (for instance, the production of $\zeta^2$ and $\zeta^3$ correlators) take place in the region $(8 \xi)^{-1} \lsim -k\tau \lsim 2\xi$ of phase space.  In this regime we can take the large argument asymptotics of the Bessel function in (\ref{bessel}) to obtain a
very simple representation of the modes:
\begin{eqnarray}
  A_{+}(\tau,k) &\cong& \frac{1}{\sqrt{2k}} \left(\frac{-k\tau}{2\xi}\right)^{1/4} e^{\pi \xi - 2\sqrt{-2\xi k\tau}} 
\nonumber\\
  A_{+}'(\tau,k) &\cong& \sqrt{\frac{2k\xi}{-\tau}} A_{+}(\tau,k)      
  \label{lorenzo}   
\end{eqnarray}

Throughout the majority of this paper we employ the representation (\ref{lorenzo}) of the modes, for brevity of exposition.  However, we have verified that none of our results  changes significantly if we use the more accurate expression (\ref{bessel}).  Formally, the only effect of using   (\ref{bessel}) rather than  (\ref{lorenzo}) is that, for any two (rescaled) momenta ${\bf q_1}$ and ${\bf q_2}$, the quantity
\begin{equation}
{\cal I} \left[ c \left( \vert {\bf q_1} \vert^{1/2} +  \vert {\bf q_2} \vert^{1/2} \right)  \right] 
\simeq \int_0^\infty d x \left( \sin x - x \cos x \right) {\rm e}^{- c \left( \vert {\bf q_1} \vert^{1/2} +  \vert {\bf q_2} \vert^{1/2} \right) \sqrt{x}}
\end{equation}
entering in eqs. (\ref{f2}) and (\ref{f3-res}) gets replaced by
\begin{eqnarray}
{\cal I}_B & \equiv &  \frac{2 c }{\pi} \, \frac{  \vert {\bf q_1} \vert^{1/4} \,  \vert {\bf q_2} \vert^{1/4} }{ 
 \vert {\bf q_1} \vert^{1/2} +  \vert {\bf q_2} \vert^{1/2} } \, \int_0^\infty d x \, \sqrt{x} \left[ \sin x - x \cos x \right]
\\
 &&\quad \times  \left[ \vert {\bf q_1} \vert^{1/2}  \, K_1 \left( c \,  \vert {\bf q_1} x  \vert^{1/2}  \right) \, 
 K_0 \left( c \,  \vert {\bf q_2} x  \vert^{1/2}  \right) +  \vert {\bf q_2} \vert^{1/2}  \, K_1 \left( c \,  \vert {\bf q_2} x  \vert^{1/2}  \right) \,  K_0 \left( c \,  \vert {\bf q_1} x  \vert^{1/2}  \right) \right]  \nonumber
\end{eqnarray}

We verified that the results obtained with this replacement are in excellent agreement with those presented in the main text. Specifically, for the nongaussianity shape we obtained identical values to those presented in Table 1 for the nongaussianity shape; for the  nongaussianity parameter  $f_{\rm NL}^{\mathrm{equil}}$ we obtained values consistent within a  few percent with those shown in Figure \ref{fig:fnlxi}. This explicitly confirms that the approximated expressions (\ref{lorenzo}) are adequate for our analysis.

\renewcommand{\theequation}{B-\arabic{equation}}
\setcounter{equation}{0}  

\section*{APPENDIX B: Perturbing the Klein-Gordon Equation}

In section \ref{sec:perts} we have derived a closed-form expression for Klein-Gordon equation (\ref{phi_metric}) for the inflaton fluctuation $\delta\varphi$ in the theory (\ref{S}).  This was derived
by employing the ADM decomposition of the metric (\ref{ADM}) and integrating out the lapse $N$ and shift $N^i$ functions.  However, one could arrive at precisely the same result by working directly
with the equations of motion.  Following \cite{malik}, we expand the metric up to second order in perturbation theory as
\begin{eqnarray}
  g_{00} &=& -a^2 (1 + 2\phi_1 + 2\phi_2) \\
  g_{0i} &=& a^2 \partial_i ( B_1 + B_2 ) \\
  g_{ij} &=& a^2 \delta_{ij}
\end{eqnarray}
Similar, we expand the scalar and gauge fields as
\begin{eqnarray}
  \varphi(t,{\bf x}) &=& \phi(t) + \delta_1\varphi(t,{\bf x}) + \delta_2 \varphi(t,{\bf x}) \\
  A_\mu(t,{\bf x}) &=& (0,\delta_1A_i(t,{\bf x}) + \delta_2 A_i(t,{\bf x}))
\end{eqnarray}

In this gauge, the curvature perturbation on uniform density hypersurfaces is $\zeta = - \frac{H}{\dot{\phi}} \, \delta \varphi$. The equation of motion for the scalar field is
\begin{equation}
\label{KGapp}
  \partial_\mu \left[ \sqrt{-g} g^{\mu\nu}\partial_\nu \varphi  \right] - \sqrt{-g} \frac{dV}{d\varphi} - \frac{\alpha}{8f}\eta^{\mu\nu\alpha\beta} F_{\mu\nu}F_{\alpha\beta} = 0
\end{equation}
We expand (\ref{KGapp}) up to second order in perturbation theory, using the Einstein constraint equations to
eliminate the metric fluctuations in order to close the system.  At linear order we find
\begin{equation}
  \left[ \partial_\tau^2 + 2\sH\partial_\tau - \grad^2 + \left( a^2 m^2  - \frac{3 \, \phi^{'2}}{M_p^2}\right) \right] \delta_1 \varphi = 0
\end{equation}
to leading order in slow roll.  At second order in perturbation theory we find  
\begin{eqnarray}
  &&  \left[ \frac{\partial^2}{\partial_\tau^2} + 2\sH\frac{\partial}{\partial_\tau} - \grad^2 + \left( a^2 m^2  - \frac{3 \, \phi^{'2}}{M_p^2}\right) \right] \delta_2 \varphi 
       = -\frac{\alpha}{f}\frac{1}{a^2} \epsilon_{ijk} \delta_1 A_i'\partial_j(\delta_1 A_k) \nonumber \\
  && + \frac{\phi'}{2 a^2 \sH M_p^2}\left[ -\frac{1}{2} \delta_1 A_i' \delta_1 A_i' -\frac{1}{4}\delta_1F_{ij} \delta_1 F_{ij} + \partial^{-2}\partial_i\left(\delta_1 F_{ij}\delta_1A_j'\right)'  \right] 
+ \mathcal{O}\left[(\delta_1\varphi)^2\right]
\end{eqnarray}
to leading order in slow roll. These results coincide with those reported in section \ref{sec:perts}, which were derived starting from the action of the perturbations. Besides being a check on our algebra, this agreement  strengthens the
conclusions of \cite{differentKG}, helping to establish the consistency of the two most popular approaches to nonlinear cosmological perturbation theory.




\bibliographystyle{apsrmp}
\bibliography{rmp-sample}

\begin{thebibliography}{99}  

\bibitem{delicate}

  D.~Baumann, A.~Dymarsky, I.~R.~Klebanov, L.~McAllister and P.~J.~Steinhardt,
  ``A Delicate Universe,''
  Phys.\ Rev.\ Lett.\  {\bf 99}, 141601 (2007)
  [arXiv:0705.3837 [hep-th]].

\bibitem{warm}

  A.~Berera,
  ``Warm Inflation,''
  Phys.\ Rev.\ Lett.\  {\bf 75}, 3218 (1995)
  [arXiv:astro-ph/9509049].

\bibitem{trapped}

 D.~Green, B.~Horn, L.~Senatore and E.~Silverstein,
  ``Trapped Inflation,''
  Phys.\ Rev.\  D {\bf 80}, 063533 (2009)
  [arXiv:0902.1006 [hep-th]].

\bibitem{DBI}

  M.~Alishahiha, E.~Silverstein and D.~Tong,
  ``DBI in the sky,''
  Phys.\ Rev.\  D {\bf 70}, 123505 (2004)
  [arXiv:hep-th/0404084].

\bibitem{NL}

  N.~Barnaby, T.~Biswas and J.~M.~Cline,
  ``p-adic inflation,''
  JHEP {\bf 0704}, 056 (2007)
  [arXiv:hep-th/0612230].
  N.~Barnaby and J.~M.~Cline,
  ``Large Nongaussianity from Nonlocal Inflation,''
  JCAP {\bf 0707}, 017 (2007)
  [arXiv:0704.3426 [hep-th]].
  N.~Barnaby and J.~M.~Cline,
  ``Predictions for Nongaussianity from Nonlocal Inflation,''
  JCAP {\bf 0806}, 030 (2008)
  [arXiv:0802.3218 [hep-th]].

\bibitem{natural}

 K.~Freese, J.~A.~Frieman and A.~V.~Olinto,
  ``Natural inflation with pseudo - Nambu-Goldstone bosons,''
  Phys.\ Rev.\ Lett.\  {\bf 65}, 3233 (1990).


\bibitem{natural1.5}

  F.~C.~Adams, J.~R.~Bond, K.~Freese, J.~A.~Frieman and A.~V.~Olinto,
  ``Natural Inflation: Particle Physics Models, Power Law Spectra For Large
  Scale Structure, And Constraints From Cobe,''
  Phys.\ Rev.\  D {\bf 47}, 426 (1993)
  [arXiv:hep-ph/9207245].

\bibitem{extranatural}

  N.~Arkani-Hamed, H.~C.~Cheng, P.~Creminelli and L.~Randall,
  ``Extranatural inflation,''
  Phys.\ Rev.\ Lett.\  {\bf 90}, 221302 (2003)
  [arXiv:hep-th/0301218].

\bibitem{2-flation}

  J.~E.~Kim, H.~P.~Nilles and M.~Peloso,
  ``Completing natural inflation,''
  JCAP {\bf 0501}, 005 (2005)
  [hep-ph/0409138].

\bibitem{N-flation}

  S.~Dimopoulos, S.~Kachru, J.~McGreevy and J.~G.~Wacker,
  ``N-flation,''
  JCAP {\bf 0808}, 003 (2008)
  [hep-th/0507205].

\bibitem{N-flation2}

  R.~Easther and L.~McAllister,
  ``Random matrices and the spectrum of N-flation,''
  JCAP {\bf 0605}, 018 (2006)
  [hep-th/0512102].

\bibitem{monodromy}

  L.~McAllister, E.~Silverstein and A.~Westphal,
  ``Gravity Waves and Linear Inflation from Axion Monodromy,''
  Phys.\ Rev.\  D {\bf 82}, 046003 (2010)
  [arXiv:0808.0706].

\bibitem{monodromy2}

  R.~Flauger, L.~McAllister, E.~Pajer, A.~Westphal and G.~Xu,
  ``Oscillations in the CMB from Axion Monodromy Inflation,''
  JCAP {\bf 1006}, 009 (2010)
  [arXiv:0907.2916].

\bibitem{kaloper}
  N.~Kaloper and L.~Sorbo,
  Phys.\ Rev.\ Lett.\  {\bf 102}, 121301 (2009)
  [arXiv:0811.1989].

\bibitem{lorenzo}

  M.~M.~Anber and L.~Sorbo,
  ``Naturally inflating on steep potentials through electromagnetic
  dissipation,''
  Phys.\ Rev.\  D {\bf 81}, 043534 (2010)
  [arXiv:0908.4089].

\bibitem{natural2}

  C.~Savage, K.~Freese and W.~H.~Kinney,
  ``Natural Inflation: status after WMAP 3-year data,''
  Phys.\ Rev.\  D {\bf 74}, 123511 (2006)
  [hep-ph/0609144].


\bibitem{big_f}

  T.~Banks, M.~Dine, P.~J.~Fox and E.~Gorbatov,
  ``On the possibility of large axion decay constants,''
  JCAP {\bf 0306}, 001 (2003)
  [hep-th/0303252].


  
\bibitem{ai}

  N.~Barnaby and M.~Peloso,
  ``Large Nongaussianity in Axion Inflation,''
  arXiv:1011.1500 [hep-ph].

\bibitem{NGreview}

  N.~Barnaby,
  ``Nongaussianity from Particle Production During Inflation,''
  Adv.\ Astron.\  {\bf 2010}, 156180 (2010)
  [arXiv:1010.5507].

\bibitem{riotto}

  V.~Acquaviva, N.~Bartolo, S.~Matarrese and A.~Riotto,
  ``Second-order cosmological perturbations from inflation,''
  Nucl.\ Phys.\  B {\bf 667}, 119 (2003)
  [arXiv:astro-ph/0209156].

\bibitem{maldacena}

  J.~M.~Maldacena,
  ``Non-Gaussian features of primordial fluctuations in single field
  inflationary models,''
  JHEP {\bf 0305}, 013 (2003)
  [arXiv:astro-ph/0210603].

\bibitem{seerylidsey}

  D.~Seery and J.~E.~Lidsey,
  ``Primordial non-gaussianities in single field inflation,''
  JCAP {\bf 0506}, 003 (2005)
  [arXiv:astro-ph/0503692].

\bibitem{small_sound}

  X.~Chen, M.~x.~Huang, S.~Kachru and G.~Shiu,
  ``Observational signatures and non-Gaussianities of general single field
  inflation,''
  JCAP {\bf 0701}, 002 (2007)
  [arXiv:hep-th/0605045].

\bibitem{nonBD1}

 P.~D.~Meerburg, J.~P.~van der Schaar and P.~S.~Corasaniti,
  ``Signatures of Initial State Modifications on Bispectrum Statistics,''
  JCAP {\bf 0905}, 018 (2009)
  [arXiv:0901.4044 [hep-th]].

\bibitem{nonBD2}

  P.~D.~Meerburg, J.~P.~van der Schaar and M.~G.~Jackson,
  ``Bispectrum signatures of a modified vacuum in single field inflation with a
  small speed of sound,''
  arXiv:0910.4986 [hep-th].

\bibitem{nonBD}

  R.~Holman and A.~J.~Tolley,
  ``Enhanced Non-Gaussianity from Excited Initial States,''
  JCAP {\bf 0805}, 001 (2008)
  [arXiv:0710.1302 [hep-th]].

\bibitem{chen1}

  X.~Chen, R.~Easther and E.~A.~Lim,
  ``Large non-Gaussianities in single field inflation,''
  JCAP {\bf 0706}, 023 (2007)
  [arXiv:astro-ph/0611645].

\bibitem{chen2}

  X.~Chen, R.~Easther and E.~A.~Lim,
  ``Generation and Characterization of Large Non-Gaussianities in Single Field
  Inflation,''
  JCAP {\bf 0804}, 010 (2008)
  [arXiv:0801.3295 [astro-ph]].

\bibitem{turnNG}

  G.~I.~Rigopoulos, E.~P.~S.~Shellard and B.~J.~W.~van Tent,
  ``Large non-Gaussianity in multiple-field inflation,''
  Phys.\ Rev.\  D {\bf 73}, 083522 (2006)
  [arXiv:astro-ph/0506704].

  F.~Vernizzi and D.~Wands,
  ``Non-Gaussianities in two-field inflation,''
  JCAP {\bf 0605}, 019 (2006)
  [arXiv:astro-ph/0603799].

 C.~T.~Byrnes, K.~Y.~Choi and L.~M.~H.~Hall,
  ``Conditions for large non-Gaussianity in two-field slow-roll inflation,''
  JCAP {\bf 0810}, 008 (2008)
  [arXiv:0807.1101 [astro-ph]].

  C.~T.~Byrnes and G.~Tasinato,
  ``Non-Gaussianity beyond slow roll in multi-field inflation,''
  JCAP {\bf 0908}, 016 (2009)
  [arXiv:0906.0767 [astro-ph.CO]].

  X.~Chen and Y.~Wang,
  ``Quasi-Single Field Inflation and Non-Gaussianities,''
  arXiv:0911.3380 [hep-th].

\bibitem{preheatNG}

  N.~Barnaby and J.~M.~Cline,
  ``Nongaussian and nonscale-invariant perturbations from tachyonic  preheating
  in hybrid inflation,''
  Phys.\ Rev.\  D {\bf 73}, 106012 (2006)
  [arXiv:astro-ph/0601481].

  N.~Barnaby and J.~M.~Cline,
  ``Nongaussianity from Tachyonic Preheating in Hybrid Inflation,''
  Phys.\ Rev.\  D {\bf 75}, 086004 (2007)
  [arXiv:astro-ph/0611750].

\bibitem{preheatNG2}

  J.~R.~Bond, A.~V.~Frolov, Z.~Huang and L.~Kofman,
  ``Non-Gaussian Spikes from Chaotic Billiards in Inflation Preheating,''
  Phys.\ Rev.\ Lett.\  {\bf 103}, 071301 (2009)
  [arXiv:0903.3407 [astro-ph.CO]].

\bibitem{pp1}

  N.~Barnaby, Z.~Huang, L.~Kofman and D.~Pogosyan,
  ``Cosmological Fluctuations from Infra-Red Cascading During Inflation,''
  Phys.\ Rev.\  D {\bf 80}, 043501 (2009)
  [arXiv:0902.0615 [hep-th]].

\bibitem{pp2}

  N.~Barnaby and Z.~Huang,
  ``Particle Production During Inflation: Observational Constraints and
  Signatures,''
  Phys.\ Rev.\  D {\bf 80}, 126018 (2009)
  [arXiv:0909.0751 [astro-ph.CO]].

\bibitem{pp3}

  N.~Barnaby,
  ``On Features and Nongaussianity from Inflationary Particle Production,''
  arXiv:1006.4615 [astro-ph.CO].

\bibitem{shandera}

  M.~LoVerde, A.~Miller, S.~Shandera {\it et al.},
  ``Effects of Scale-Dependent Non-Gaussianity on Cosmological Structures,''
  JCAP {\bf 0804}, 014 (2008).
  [arXiv:0711.4126 [astro-ph]].

\bibitem{Sorbo:2011rz}
  L.~Sorbo,
  ``Parity violation in the Cosmic Microwave Background from a pseudoscalar
  inflaton,''
  arXiv:1101.1525 [astro-ph.CO].


\bibitem{Saito:2007kt}
  S.~Saito, K.~Ichiki, A.~Taruya,
  ``Probing polarization states of primordial gravitational waves with CMB anisotropies,''
  JCAP {\bf 0709}, 002 (2007).
  [arXiv:0705.3701 [astro-ph]].

\bibitem{Gluscevic:2010vv}
  V.~Gluscevic, M.~Kamionkowski,
  ``Testing Parity-Violating Mechanisms with Cosmic Microwave Background Experiments,''
  Phys.\ Rev.\  {\bf D81}, 123529 (2010).
  [arXiv:1002.1308 [astro-ph.CO]].


\bibitem{wmap7}

  E.~Komatsu {\it et al.}  [WMAP Collaboration],
  ``Seven-Year Wilkinson Microwave Anisotropy Probe (WMAP) Observations:
  Cosmological Interpretation,''
  arXiv:1001.4538 [astro-ph.CO].


\bibitem{seery}

  D.~Seery,
  ``Magnetogenesis and the primordial non-gaussianity,''
  JCAP {\bf 0908}, 018 (2009)
  [arXiv:0810.1617 [astro-ph]].

\bibitem{seerymulti}

  D.~Seery and J.~E.~Lidsey,
  ``Primordial non-gaussianities from multiple-field inflation,''
  JCAP {\bf 0509}, 011 (2005)
  [arXiv:astro-ph/0506056].

\bibitem{differentKG}

  K.~A.~Malik, D.~Seery and K.~N.~Ananda,
  ``Different approaches to the second order Klein-Gordon equation,''
  Class.\ Quant.\ Grav.\  {\bf 25}, 175008 (2008)
  [arXiv:0712.1787 [astro-ph]].

\bibitem{KGNG}

  D.~Seery, K.~A.~Malik and D.~H.~Lyth,
  ``Non-gaussianity of inflationary field perturbations from the field equation,''
  JCAP {\bf 0803}, 014 (2008)
  [arXiv:0802.0588 [astro-ph]].

\bibitem{leblond}

  L.~Leblond and E.~Pajer,
  ``Resonant Trispectrum and a Dozen More Primordial N-point functions,''
  arXiv:1010.4565 [hep-th].

\bibitem{malik}

  K.~A.~Malik,
  ``A not so short note on the Klein-Gordon equation at second order,''
  JCAP {\bf 0703}, 004 (2007)
  [arXiv:astro-ph/0610864].

\bibitem{dN}

  D.~H.~Lyth, K.~A.~Malik and M.~Sasaki,
  ``A general proof of the conservation of the curvature perturbation,''
  JCAP {\bf 0505}, 004 (2005)
  [arXiv:astro-ph/0411220].

\bibitem{shape}

  D.~Babich, P.~Creminelli and M.~Zaldarriaga,
  ``The shape of non-Gaussianities,''
  JCAP {\bf 0408}, 009 (2004)
  [arXiv:astro-ph/0405356].

\bibitem{fnlbounds}

  L.~Senatore, K.~M.~Smith and M.~Zaldarriaga,
  ``Non-Gaussianities in Single Field Inflation and their Optimal Limits from
  the WMAP 5-year Data,''
  JCAP {\bf 1001}, 028 (2010)
  [arXiv:0905.3746 [astro-ph.CO]].


\bibitem{kallosh}

  R.~Kallosh,
  ``On Inflation in String Theory,''
  Lect.\ Notes Phys.\  {\bf 738}, 119 (2008)
  [arXiv:hep-th/0702059].

\bibitem{bond}

  F.~C.~Adams, J.~R.~Bond, K.~Freese {\it et al.},
  ``Natural inflation: Particle physics models, power law spectra for large scale structure, and constraints from COBE,''
  Phys.\ Rev.\  {\bf D47}, 426-455 (1993).
  [hep-ph/9207245].

\bibitem{assisted}

 A.~R.~Liddle, A.~Mazumdar and F.~E.~Schunck,
 ``Assisted inflation,''
  Phys.\ Rev.\  D {\bf 58}, 061301 (1998)
  [arXiv:astro-ph/9804177].

\bibitem{kallosh2}

  R.~Kallosh, N.~Sivanandam and M.~Soroush,
  ``Axion Inflation and Gravity Waves in String Theory,''
  Phys.\ Rev.\  D {\bf 77}, 043501 (2008)
  [arXiv:0710.3429 [hep-th]].

\bibitem{bm}

  D.~Baumann and L.~McAllister,
  ``Advances in Inflation in String Theory,''
  Ann.\ Rev.\ Nucl.\ Part.\ Sci.\  {\bf 59}, 67 (2009)
  [arXiv:0901.0265 [hep-th]].

\bibitem{N105}

  P.~Candelas, E.~Perevalov and G.~Rajesh,
  ``Toric geometry and enhanced gauge symmetry of F-theory/heterotic  vacua,''
  Nucl.\ Phys.\  B {\bf 507}, 445 (1997)
  [arXiv:hep-th/9704097].

\bibitem{grimm}

  T.~W.~Grimm,
  ``Axion inflation in type II string theory,''
  Phys.\ Rev.\  {\bf D77}, 126007 (2008).
  [arXiv:0710.3883 [hep-th]].

\bibitem{green}

  D.~R.~Green,
  ``Reheating Closed String Inflation,''
  Phys.\ Rev.\  D {\bf 76}, 103504 (2007)
  [arXiv:0707.3832 [hep-th]].


\bibitem{multi}

  J.~Braden, L.~Kofman and N.~Barnaby,
  ``Reheating the Universe After Multi-Field Inflation,''
  JCAP {\bf 1007}, 016 (2010)
  [arXiv:1005.2196 [hep-th]].

\bibitem{Nmag}

  M.~M.~Anber and L.~Sorbo,
  ``N-flationary magnetic fields,''
  JCAP {\bf 0610}, 018 (2006)
  [arXiv:astro-ph/0606534].


\bibitem{monodromyNG}

  S.~Hannestad, T.~Haugbolle, P.~R.~Jarnhus and M.~S.~Sloth,
  ``Non-Gaussianity from Axion Monodromy Inflation,''
  JCAP {\bf 1006}, 001 (2010)
  [arXiv:0912.3527 [hep-ph]].

\bibitem{pajer}

 R.~Flauger and E.~Pajer,
  ``Resonant Non-Gaussianity,''
  arXiv:1002.0833 [hep-th].


\bibitem{mixing2}

  N.~Kaloper, A.~Lawrence and L.~Sorbo,
  ``An Ignoble Approach to Large Field Inflation,''
  arXiv:1101.0026 [hep-th].

\bibitem{dante}

  M.~Berg, E.~Pajer and S.~Sjors,
  ``Dante's Inferno,''
  Phys.\ Rev.\  D {\bf 81}, 103535 (2010)
  [arXiv:0912.1341 [hep-th]].

\bibitem{racetrack1}

 J.~J.~Blanco-Pillado {\it et al.},
  ``Racetrack inflation,''
  JHEP {\bf 0411}, 063 (2004)
  [arXiv:hep-th/0406230].

\bibitem{racetrack2}

  J.~J.~Blanco-Pillado {\it et al.},
  ``Inflating in a better racetrack,''
  JHEP {\bf 0609}, 002 (2006)
  [arXiv:hep-th/0603129].

\bibitem{D3D7}

  C.~P.~Burgess, J.~M.~Cline, M.~Postma,
  ``Axionic D3-D7 Inflation,''
  JHEP {\bf 0903}, 058 (2009).
  [arXiv:0811.1503 [hep-th]].


\bibitem{LVC1}

  V.~Balasubramanian and P.~Berglund,
  ``Stringy corrections to Kahler potentials, SUSY breaking, and the
  cosmological constant problem,''
  JHEP {\bf 0411}, 085 (2004)
  [arXiv:hep-th/0408054].

\bibitem{LVC2}

  V.~Balasubramanian, P.~Berglund, J.~P.~Conlon and F.~Quevedo,
  ``Systematics of Moduli Stabilisation in Calabi-Yau Flux Compactifications,''
  JHEP {\bf 0503}, 007 (2005)
  [arXiv:hep-th/0502058].

\bibitem{LVC3}

 J.~P.~Conlon, F.~Quevedo and K.~Suruliz,
  ``Large-volume flux compactifications: Moduli spectrum and D3/D7 soft
  supersymmetry breaking,''
  JHEP {\bf 0508}, 007 (2005)
  [arXiv:hep-th/0505076].

\bibitem{roulette}

  J.~R.~Bond, L.~Kofman, S.~Prokushkin and P.~M.~Vaudrevange,
  ``Roulette inflation with Kaehler moduli and their axions,''
  Phys.\ Rev.\  D {\bf 75}, 123511 (2007)
  [arXiv:hep-th/0612197].

\bibitem{CQ}

  J.~P.~Conlon and F.~Quevedo,
  ``Kaehler moduli inflation,''
  JHEP {\bf 0601}, 146 (2006)
  [arXiv:hep-th/0509012].

\bibitem{astro}

  J.~P.~Conlon and F.~Quevedo,
  ``Astrophysical and Cosmological Implications of Large Volume String
  Compactifications,''  
  JCAP {\bf 0708}, 019 (2007)
  [arXiv:0705.3460 [hep-ph]].

\bibitem{modular}

  N.~Barnaby, J.~R.~Bond, Z.~Huang and L.~Kofman,
  ``Preheating After Modular Inflation,''
  JCAP {\bf 0912}, 021 (2009)
  [arXiv:0909.0503 [hep-th]].

\bibitem{misra}

  A.~Misra, P.~Shukla,
  ``Large Volume Axionic Swiss-Cheese Inflation,''
  Nucl.\ Phys.\  {\bf B800}, 384-400 (2008).
  [arXiv:0712.1260 [hep-th]].


\bibitem{pdg}
K. Nakamura et al. (Particle Data Group), J. Phys. G 37, 075021 (2010).



\bibitem{abramowitz}
M. Abramowitz and I. A. Stegun ,
``Handbook of Mathematical Functions''













\end{thebibliography}

\end{document}